\documentclass[a4paper,11pt]{article}
\pdfoutput=1 % if your are submitting a pdflatex (i.e. if you have
\usepackage{jheppub}
\usepackage{xcolor}% for details on the use of the package, please
\definecolor{darkgreen}{rgb}{0.42, 0.70, 0.13}

\usepackage[T1]{fontenc} % if needed
\usepackage{bbold}
\usepackage{amsmath}
\usepackage{amssymb}
\usepackage{mathtools}
\usepackage{stmaryrd}
\usepackage{tensor}
\usepackage{multirow}
\usepackage{multicol}
\usepackage{bigstrut}

\usepackage{subcaption}
\usepackage{graphicx}
\usepackage{stmaryrd}
\let\a=\alpha 
\let\b=\beta 
 
\let\d=\delta 
\let\e=\epsilon

\let\q=\theta

\let\r=\rho
 
\let\t=\tau  
\let\f=\phi

\let\y=\psi

\let\W=\Omega   
\let\G=\Gamma

\let\del=\partial
\let\<=\langle
\let\>=\rangle

\newcommand{\CC}{\mathcal{C}}

\newcommand{\GG}{\mathcal{G}}

\newcommand{\NN}{\mathcal{N}}

\def\be#1\ee{\begin{align}#1\end{align}}

\newcommand{\op}[1]{\operatorname{#1}}
\newcommand{\al}[1]{\begin{align}#1\end{align}}
\newcommand{\spl}[1]{\begin{split}#1\end{split}}
\newcommand{\mat}[1]{\begin{pmatrix}#1\end{pmatrix}}

\newcommand{\circled}[1]{\raisebox{.5pt}{\textcircled{\raisebox{-.9pt} {#1}}}}

\newcommand{\XZ}[1]{{{\color[rgb]{1,1,1}{}}}}
\newcommand{\JP}[1]{\textbf{\color[rgb]{1,1,1}{}}}
\newcommand{\SK}[1]{\textbf{\color[rgb]{1,1,1}{}}}
\newcommand{\AM}[1]{\textbf{\color[rgb]{1,1,1}{}}}
\newcommand{\AV}[1]{\textbf{\color[rgb]{1,1,1}{}}}

\usepackage{tikz}
\tikzset{every picture/.style={line width=0.8pt}}
\tikzset{graph-1/.style = {
  line cap = round,
   line join = round,
     > = triangle 45,
     x=0.7cm, y=0.7cm,
      every node/.append style = {inner ysep=2mm}
                        }
    }% end of tikzset
\usetikzlibrary{matrix,decorations.pathreplacing}

\title{
Einstein gravity from a matrix integral - Part I
}

\author[a]{Shota Komatsu}
\author[b]{Adrien Martina}
\author[b]{Joao Penedones}
\author[b]{Antoine Vuignier}
\author[b,c]{Xiang Zhao}
\affiliation[a]{CERN, Theoretical Physics Department,
CH-1211 Geneva 23, Switzerland}
\affiliation[b]{Fields and Strings Laboratory, Institute of Physics, Ecole Polytechnique Federale de Lausanne (EPFL),
CH-1015 Lausanne, Switzerland}
\affiliation[c]{Université Paris-Saclay, CNRS, CEA, Institut de Physique Théorique, 91191, Gif-sur-Yvette, France
}

\abstract{
We construct 
backreacted geometries dual to the supersymmetric mass deformation of the IKKT matrix model. They are Euclidean type IIB supergravity solutions given in terms of an electrostatic potential, having $SO(7)\times SO(3)$ isometry and 16 supersymmetries. Quantizing the fluxes, we find that the supergravity solutions are in one-to-one correspondence with fuzzy sphere vacua of the matrix model. 
\\
}

\begin{document} 
\maketitle
\flushbottom

\newpage
\section{Introduction}

The gauge/gravity duality is our best  UV complete model of quantum gravity.
It tells us that some quantum mechanical systems are well described by Einstein gravity in some regimes, usually involving many strongly coupled degrees of freedom.
We would like to find out what is special about such systems. 

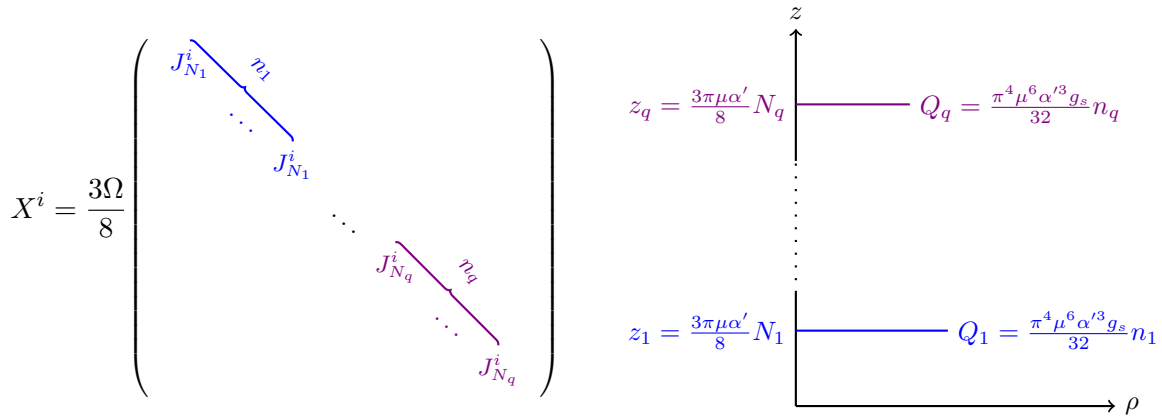
\begin{figure}[h]
\[ 
   X^i = \frac{3 \Omega}{8}
   \begin{tikzpicture}[decoration={brace,amplitude=2pt},baseline]
   \footnotesize
   \matrix (magic) [matrix of math nodes,left delimiter=(,right delimiter=)] {
      \color{blue}J^i_{N_1}  &  &  \\
     & \color{blue} \ddots  &  \\
     &  &  \color{blue} J^i_{N_1} \\
     & & & \ddots \\
     & & & & \color{violet} J^i_{N_q} \\
     & & & & & \color{violet} \ddots \\
     & & & & & & \color{violet} J^i_{N_q}\\
    };
    \draw[decorate,blue] (magic-1-1.north) -- (magic-3-3.north) node[above=5pt,midway,sloped] {$n_1$};
    \draw[decorate,violet] (magic-5-5.north) -- (magic-7-7.north) node[above=5pt,midway,sloped] {$n_q$};
  \end{tikzpicture}
 \qquad 
 \begin{tikzpicture}[baseline] 
\small
  \draw[->] (0,-2.5) -- (4.2,-2.5) node[right] {$\rho$};
  \draw (0,-2.5) -- (0,-1) ;
   \draw[loosely dotted] (0,-1) -- (0,0.75) ;
  \draw[->] (0,0.75) -- (0,2.5) node[above] {$z$};
  \draw[color=blue]   (0,-1.5)  node[left]{$z_1 = \frac{3\pi \mu\alpha'}{8} N_1$} -- (2,-1.5) node[right]{$Q_1 = \frac{\pi^4 \mu^6\alpha'^3 g_s}{32} n_1$};
  \draw[color=violet]   (0,1.5) node[left]{$z_q = \frac{3\pi \mu \alpha'}{8} N_q$} -- (1.5,1.5) node[right]{$Q_q = \frac{\pi^4 \mu^6\alpha'^3 g_s}{32} n_q$};

\end{tikzpicture}
\]
  \caption{Correspondence between the fuzzy sphere vacua of the matrix model and the dual geometries. For each spin $j_s$ $SU(2)$ irreducible representation of dimension $N_s=2j_s+1$ we put a disk at position $z_s \sim N_s$. The number $n_s$ of copies of this representation appearing determines the charge of that disk $Q_s \sim n_s$. Those integers are related to the $D1$ brane charge and $NS5$ brane charge  by $ N_{D1,s}=n_s$ and $N_{NS5,s} = N_s-N_{s-1}$. }
  \label{fig:vacua}
\end{figure}

In this paper, we study the simplest toy model that has an emergent Einstein gravity description, namely, a matrix integral. More precisely, we 
study the {\it polarized IKKT matrix model} \cite{ Bonelli_2002, Hartnoll:2024csr}. 
This is a supersymmetric mass deformation of the Ishibashi-Kawai-Kitazawa-Tsuchiya (IKKT) matrix model \cite{Ishibashi_1997}.
It has 
vacua\footnote{By vacuum we simply mean a local minimum of the action.} given by $N$-dimensional representations of $SU(2)$. These fuzzy sphere configurations are depicted on the left of Figure \ref{fig:vacua}.
For $N \times N$ matrices, the number of such vacua is $P(N)$, the number of integer partitions of $N$. 
%Those vacua are non degenerate and the dominant contribution comes from the largest irreducible representation. We also show that those vacua preserve \JP{half of the} supersymmetries of the model. \XZ{Is ``half of the susy's'' = sixteen susy's?}

%\SK{Added} 
The IKKT matrix model was originally proposed as a non-perturbative definition of (type IIB) string theory. This is similar in spirit to the BFSS conjecture \cite{Banks:1996vh,Susskind:1997cw}, which proposes that the BFSS matrix quantum mechanics provides a non-perturbative definition of M-theory in flat space. While testing or verifying the BFSS conjecture remains challenging (see \cite{Herderschee:2023pza, Komatsu:2024vnb} and references therein for recent discussions), the BFSS model was later revisited from the perspective of standard holography %\XZ{added \cite{Polchinski:1999br}}
\cite{Itzhaki_1998,Polchinski:1999br}, which helped to clarify its relation to M-theory and sharpen the original conjecture. In contrast, little work has been done on holography for the IKKT model, with the notable exception of studies of the decoupling limit of the D-instanton background \cite{Ooguri:1998pf,Gibbons:1995vg,Bergshoeff:1998ry}. This was partly due to the absence of observables that could be computed and compared on both sides. In this paper and a companion paper \cite{Komatsu:2024ydh}, we demonstrate a one-to-one correspondence between vacua of the polarized IKKT matrix integral and the dual backreacted geometries.

More precisely, we identify a family of geometries dual to the polarized IKKT matrix integral, that are solutions of Euclidean type IIB supergravity with vanishing 5-form flux, see \eqref{eq:sugra_solution}. The metric takes the form of a warped product, with a 2-sphere  and a 6-sphere fibered over a 2-dimensional plane parametrized by $(\rho,z)$,% The metric takes the form
\begin{equation}
    ds^2 = R_2(\rho,z)^2 d\Omega_2^2 + R_6(\rho,z)^2 d\Omega_6^2 + H(\rho,z)^2 (d\rho^2+dz^2).
\end{equation}
The solutions are given in terms of a single function $V(\rho,z)$ that solves the 4 dimensional axially symmetric Laplace equation,
\begin{equation}
    V'' + \frac{2}{\rho}\dot{V} + \ddot{V}=0
\end{equation}
where $V' \equiv \partial_z V$ and $\dot{V}\equiv \partial_\rho V$. We can think of $\rho$ as the radial coordinate on $\mathbb{R}^3$ and $z$ a coordinate along the orthogonal direction in $\mathbb{R}^4$. This allows us to use an electrostatic analogy, where $V$ is an electrostatic potential. This is  similar to the Lin-Maldacena geometries \cite{Lin:2005nh,Lin:2004nb} that describe the vacua of the BMN model \cite{Berenstein:2002jq}.

Let us describe the electrostatic analogy in more detail. Boundary conditions are imposed by considering $q$ conducting disks\footnote{We use the word \emph{disks} by analogy with \cite{Lin:2005nh} but in our case we have 3D conducting balls in $\mathbb{R}^4$.} of radius $\rho_s$, centered on the $z$-axis at different positions $z_s$ and of given charges $Q_s$, with $s=1,...,q$. In addition we add a background potential that grows at infinity,
%\AV{added eqref} 
see \eqref{eq: Vbg}, and an infinite grounded disk (or grounded plane) at $z=0$. Those boundary conditions come from requiring that the supergravity solution is regular. 
% \JP{REMOVE:, which means that the metric and dilaton are positive, and all the supergravity fields don't have any singularity.} 
This also determines the size of the disks $\rho_s(Q_s,z_s)$ 
%\AM{Did you define $d_i$ ? Is it $z_{i + 1} - z_i$ ?} 
since in general we expect the electric field to be infinite at the tip of a disk. However, since there is a background potential, we can choose the size of the disks so that the total electric field vanishes at the tip. This ensures that we don't have any singularity there.

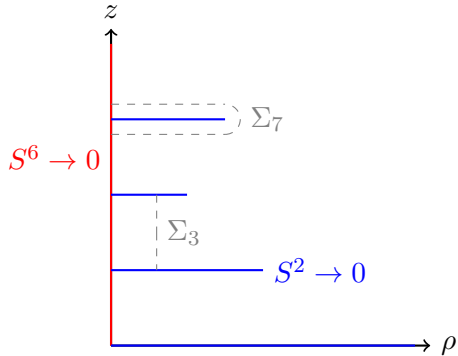
\begin{figure}[t]
    \centering
    \begin{tikzpicture}[domain=0:4]

  \draw[->] (0,0) -- (4.2,0) node[right] {$\rho$};
  \draw[->] (0,0) -- (0,4.2) node[above] {$z$};

  \draw[color=red]    (0,0) -- (0,2.5) node[left]{$S^6 \to 0$} (0,2.5) -- (0,4);
  \draw[color=blue]   (0,0) -- (4,0);
  \draw[color=blue]   (0,1) -- (2,1) node[right]{$S^2 \to 0$};
  \draw[color=blue]   (0,2) -- (1,2);
  \draw[color=blue]   (0,3) -- (1.5,3);
  \draw[thin, dashed,color=gray]   (0,3.2) -- (1.5,3.2);
  \draw[thin, dashed,color=gray]   (0,2.8) -- (1.5,2.8);
\draw[thin, dashed,color=gray] (1.5,2.8) arc(-90:0:0.2) node[right]{$\Sigma_7$} arc(0:90:0.2);
\draw[thin, dashed,color=gray] (0.6,2) -- (0.6,1.5)node[right]{$\Sigma_3$} (0.6,1.5)--(0.6,1) ;
\end{tikzpicture}
    \caption{Picture of the geometry in the $(\rho,z)$ plane. The 10d geometry is obtained by fibering an $S^2$ and an $S^6$. The blue lines are defined by $\dot{V}\equiv \partial_\rho V=0$ and are the regions where $S^2$ shrinks. The red line is the $z$-axis $\rho=0$ where the $S^6$ shrinks. We can construct 7-cycles $\Sigma_7$ as the product of the 6-sphere times a segment on the $(\rho,z)$ plane whose endpoints have $\rho=0$ (where $S^6$ shrinks). Similarly, there are 3-cycles $\Sigma_3$ given by the 2-sphere times a segment connecting points where $S^2$ shrinks.}
    \label{fig:disks_plane}
\end{figure}

These geometries are topologically non-trivial and they have various non-contractible cycles. This comes from the fact that the 6-sphere collapses %$S^6 \to 0$ 
on the $z$-axis and the 2-sphere collapses %$S^2\to 0$
on the disks. We can then construct 3-cycles and 7-cycles by considering curves in the $(\rho,z)$ plane ending on locations where the spheres shrink. The cycles are constructed by fibering a sphere over such curves, as shown in Figure \ref{fig:disks_plane}.  

% \begin{figure}[h]
%     \centering
%     \begin{tikzpicture}[domain=0:4]

%   \draw[->] (0,0) -- (4.2,0) node[right] {$\rho$};
%   \draw[->] (0,0) -- (0,4.2) node[above] {$z$};

%   \draw[color=red]    (0,0) -- (0,2.5) node[left]{$S^6 \to 0$} (0,2.5) -- (0,4);
%   \draw[color=blue]   (0,0) -- (4,0);
%   \draw[color=blue]   (0,1) -- (2,1) node[right]{$S^2 \to 0$};
%   \draw[color=blue]   (0,2) -- (1,2);
%   \draw[color=blue]   (0,3) -- (1.5,3);
%   \draw[thin, dashed,color=gray]   (0,3.2) -- (1.5,3.2);
%   \draw[thin, dashed,color=gray]   (0,2.8) -- (1.5,2.8);
% \draw[thin, dashed,color=gray] (1.5,2.8) arc(-90:0:0.2) node[right]{$\Sigma_7$} arc(0:90:0.2);
% \draw[thin, dashed,color=gray] (0.6,2) -- (0.6,1.5)node[right]{$\Sigma_3$} (0.6,1.5)--(0.6,1) ;
% \end{tikzpicture}
%     \caption{Picture of the geometry in the $(\rho,z)$ plane. The 10d geometry is obtained by fibering an $S^2$ and an $S^6$. The blue lines are defined by $\dot{V}\equiv \partial_\rho V=0$ and are the regions where $S^2$ shrinks. The red line is the $z$-axis $\rho=0$ where the $S^6$ shrinks. We can construct 7-cycles $\Sigma_7$ as the product of the 6-sphere times a segment on the $(\rho,z)$ plane whose endpoints have $\rho=0$ (where $S^6$ shrinks). Similarly, there are 3-cycles $\Sigma_3$ given by the 2-sphere times a segment connecting points where $S^2$ shrinks.}
%     \label{fig:disks_plane}
% \end{figure}

The different supergravity fluxes can then be integrated over those cycles. Imposing the Dirac quantization conditions we find that the charges of the disks and their positions are quantized. Relating those quantum numbers to the dimensions and degeneracies of the $SU(2)$ irreducible representations, we find a one-to-one correspondence between the fuzzy sphere vacua and the supergravity solutions, as depicted in Figure \ref{fig:vacua}.

We also study the system from the  perspective of a probe brane. We consider a probe spherical $D1$-brane in a supergravity background that has a constant Ramond-Ramond flux, and compute its size by minimizing the DBI action. We get an exact correspondence between this radius and the radius of the fuzzy sphere vacua of the matrix model. Furthermore we show that this radius can also be computed in the backreacted geometry as the geodesic distance between the grounded conducting plane and a disk.

%\JP{Paragraph to fix depending on finding the missing factors...}
%We can then compare the on-shell action of the supergravity solutions with the free energy of the mass-deformed IKKT model, also called the polarized IKKT model \cite{Hartnoll:2024csr}. Using that the on-shell Lagrangian is a total derivative, we can use Stokes theorem to compute the on-shell action. The contributions from the non-contractible cycles in the geometry can be evaluated explicitly and we get the formula \eqref{eq:free_energy} for the supergravity side. 
%We compare this result with value of the polarized IKKT action evaluated on the fuzzy sphere saddles. We find agreement \AV{$\times 4$} in the regime of large $SU(2)$ irreducible representations, where the contributions from fluctuations are suppressed.
%In upcoming work, we will use susy localization to compute these contributions and extend the matching with supergravity at large $N$.

%In principle, the free energy of the matrix  model receives more contributions coming from fluctuations around those saddles, and that can be computed via localization. We leave this computation for further work and, in the meantime, note that we exactly match the value of the polarized IKKT action evaluated on the fuzzy spheres for large representations.

\paragraph{Structure of the paper.}
In section \ref{sec:IKKT}, we review the supersymmetric mass deformation of the IKKT matrix model. We study the saddle points of the matrix integral and show that they are  fuzzy spheres.
In section \ref{sec:geometries}, we construct Euclidean type IIB geometries with $SO(3) \times SO(7)$ symmetry and 16 supercharges. 
%We formulate the solution in terms of a 4 dimensional electrostatic potential, and analyze the regularity conditions to ensure smoothness of the geometry. We study the asymptotic form of the solution where we recover 10D flat Euclidean space. We show that quantizing the different fluxes, we get a discrete set of solutions that are in one-to-one correspondence with the fuzzy sphere saddles of the mass-deformed IKKT integral. We then compute the on-shell action of the supergravity solution. We also compute the size of a probe $D1$ brane and match its radius with the size of the fuzzy spheres and the height of the disk. 
We conclude in section \ref{sec:discussion} with a discussion of our results. 
%\AV{added description of appendices:} 
Our appendix \ref{app_analytic continuation} shows how to obtain the Euclidean IIB solution from the Lorentzian of \cite{DHoker:2016ujz}. In appendix \ref{app_regularity} we show that the geometries we obtained are regular. In appendix \ref{app:electrostatics} we review the solution to the electrostatic problem when the disks are widely separated.
%\JP{there a lot of repetition of the previous paragraphs.}

\section{The IKKT model and its mass deformation}
\label{sec:IKKT}
The (Euclidean) IKKT integral 
% \AV{add footnote for fermion components} \cite{Ishibashi_1997,Aoki:1998bq} 
is 
\begin{equation}
    Z_{\rm IKKT} = \int \prod_{I=1}^{10} dX_I \prod_{\alpha=1}^{16} d\psi_\alpha \ e^{-S_{\rm IKKT}},
    \label{eq:IKKT partition function}
\end{equation}
% \AM{Updated $\prod_{\alpha=1}^{32} \to \prod_{\alpha=1}^{16}$}
where\footnote{The action has the same convention as \cite{Green:1997tn,Moore:1998et,Green:1998yf}, which is natural from the perspective of dimensionally reducing 10D $\NN=1$ SYM to a point. This convention is related to that in \cite{Hartnoll:2024csr} through a simple redefinition of the charge conjugation matrix: $i\CC_{\rm here} = \CC_{\rm there}$. After Weyl projection to the suitable 16-component spinors, the actions are the same. The Weyl projection also explains why $\a$ only goes up to 16 in the fermion integral in \eqref{eq:IKKT partition function}. For more details, see \cite{Komatsu:2024ydh}.}
\al{\spl{
    S_{\rm IKKT} &= - \operatorname{Tr} \left[\frac{1}{4} \sum_{I,J=1}^{10}[X_I,X_J]^2 + \frac{i}{2} \sum_{\alpha,\beta=1}^{32} \sum_{I=1}^{10} \psi_\alpha (\mathcal{C}\G^{I})_{\alpha \beta} [X_I,\psi_\beta] \right],
\label{eq:IKKT action}
}}
% \AM{If we want to expand all fermionic components, it would be nicer to use the notation $ S \supset - \sum_{\alpha=1}^{16} \psi_\alpha \bar \gamma_{\alpha \beta} \psi_\beta$, or I would suggest $\bar \psi \Gamma^I \psi = \psi^\top C \Gamma^I \psi$, but here it's confusing because it suggests there are 32 independent fermionic components.}\XZ{Here the goal is to introduce the IKKT model with minimal effort. The most natural thing to me is to connect to expressions in the literature. I agree the 16-component is the best for practical purposes. I have in mind of explaining the Weyl projection and switching to 16-component fermion in the susy localisation paper. It is also helpful for younger readers to be a bit pedagogical.}
where $X_I$ are $N \times N$ traceless and hermitian bosonic matrices, $\psi_\alpha$ are $N\times N$ hermitian traceless matrices of 10D Euclidean Majorana-Weyl spinors,
% \footnote{When Wick rotating from 9+1 dimensions to 10+0 dimensions, for Majorana fermions we need to double the fermionic degrees of freedom by treating $\bar\y$ as independent from $\y$. This fermion-doubling problem is resolved by working only with $\y$ but not its hermitian conjuate. See e.g. \cite[Section 2]{Nicolai:1978vc} and \cite[Section 3.1]{Yee:2003ge}. \AM{Are you sure we need these complications coming from Wick rotation ? I don't see why, starting from everything Euclidean, there would be any need for such argument. I would say we have real grassmann numbers $\psi = (\psi_{\alpha = 1,..., 16}, 0,...,0)$ and everything makes perfect sense. Maybe one should emphasize this before making comparison to the Minkowski case}\XZ{I agree from a practical perspective, especially because here we are dealing with matrices of numbers in 0+0 dimension. We can comment out this footnote here, but I don't the conceptual aspect of this comment is meaningless.}} 
and $\G^I$ are the $SO(10)$ gamma matrices, with $\mathcal{C}$ the charge conjugation matrix.\footnote{See appendix A of \cite{Pestun:2007rz} for explicit expressions of the gamma matrices. For the charge conjugation matrix we choose $\CC=\mat{0 & -i\mathbb{1} \\ i \mathbb{1} & 0}$.}

The result of the matrix integral \eqref{eq:IKKT partition function} was conjectured in \cite{Green:1997tn} and later computed in \cite{Moore:1998et} up to a group-theoretical prefactor, which was calculated in \cite{Krauth:1998xh} (see \cite{austing2001yangmillsmatrixtheory} for a review). The convergence of the partition function has been shown in \cite{Austing:2001bd,Austing:2001pk}, where they also show that the correlation functions of matrix polynomials below degree 14 are finite (see \cite{Krauth:1999rc} for some explicit correlation function examples). The gravitational dual to the IKKT model was studied in \cite{Ooguri:1998pf} by extending Maldacena's decoupling limit \cite{Maldacena:1997re,Itzhaki_1998} to $D$-instantons, and the result is flat space with non-vanishing, running dilaton and axion. In this paper we focus on Euclidean signature. For recent discussions on the Lorentzian IKKT model, see e.g. \cite{Asano:2024edo,Asano:2024def}.

The IKKT matrix model admits a mass deformation that preserves the sixteen dynamical\footnote{The IKKT model of $U(N)$ matrices has sixteen dynamical and sixteen kinematical supersymmetries \cite{Ishibashi_1997,Aoki:1998bq}, with the latter describing the center-of-mass degrees of freedom of the $D$-instantons \cite{Ishibashi_1997,Green:1997tn}. Since we choose to focus on traceless matrices and so the other sixteen kinematical supersymmetries of the IKKT model are irrelevant here.} supersymmetries \cite[App.A]{Bonelli_2002}. The action of the mass-deformed IKKT model reads
\begin{equation}
\label{mass-deformed-IKKT}
\begin{split}
    S_\Omega &= S_{\rm IKKT} + S_{\rm def}
    %- \operatorname{Tr} \left[\frac{1}{4} \sum_{I,J=1}^{10}[X_I,X_J]^2 + \frac{1}{2} \sum_{\alpha,\beta=1}^{16} \sum_{I=1}^{10} \psi_\alpha \mathcal{C}\gamma_{\alpha \beta}^{I} [X_I,\psi_\beta] \right] \\ 
    \\
    S_{\rm def}&=  \operatorname{Tr} \left[ \frac{3 \Omega^2}{4^3} \sum_{i=1}^3 X_i X_i + \frac{\Omega^2}{4^3} \sum_{p=4}^{10} X_pX_p +i \frac{\Omega}{3} \epsilon_{ijk} X_i X_j X_k - \frac{ \Omega}{8}  \psi_\a (\mathcal{C} \G^{123})_{\a\b} \psi_\b \right],
\end{split}
\end{equation}
where the indices $i,j,k \in \{1,2,3\}$, $p \in \{4,5,\dots,10\}$ and $\a,\b\in\{1,2,\dots,32\}$. Here $\Gamma^{123}=\G^1\G^2\G^3$.
Following \cite{Hartnoll:2024csr}, we will refer to this model as the \emph{polarized IKKT matrix model}. We will also take $X_I$, $\y_\a$ and $\W$ to be all dimensionless. In the limit $\W\to\infty$, $S_\W$ is dominated by the Gaussian mass terms and so the system is almost free. On the other hand, $\W\to0$ corresponds to the strong coupling limit of the polarized IKKT model.

\subsection{Symmetries} 
The mass deformation terms in \eqref{mass-deformed-IKKT} explicitly break the $SO(10)$ symmetry of the original IKKT model to $SO(3) \times SO(7)$. The matrices $X^I$ and $\psi$ transform in the adjoint representation of $SU(N)$. This is a global symmetry of the matrix model but we will often refer to  it as the \emph{gauge} symmetry by analogy with higher dimensional models. In addition, there are 16 supersymmetries acting as \begin{equation}
\begin{cases}
\delta X^I = - \y_\alpha (\mathcal{C}\G^I)_{\alpha \beta} \e_\beta\,, \\
% \delta \psi = - \frac{1}{4} [X_I, X_J] \G^{I J} \epsilon + i \frac{3 \Omega}{16} X_i \G^{1 2 3} \G^i \epsilon + i \frac{\Omega}{16} X_p \G^{1 2 3} \G^p \epsilon \,,\\
\delta \psi_\a = \frac{i}{2} [X_I, X_J] \Gamma^{I J}_{\a\b} \epsilon_\b  + \frac{3 \Omega}{8}X_i (\Gamma^{1 2 3} \Gamma^i)_{\a\b} \epsilon_\b  + \frac{\Omega}{8}X_p (\Gamma^{1 2 3} \Gamma^p)_{\a\b} \epsilon_\b \,,
\end{cases}
\label{susy_mass_deform}
\end{equation}
where $\epsilon$ is deemed as Grassmann-even and $\d$ is Grassman-odd susy generator.

%\SK{Added}
These bosonic and fermionic symmetries are expected to give (a real form of) the exceptional Lie superalgebra $F_4$, which is also the superconformal algebra in five dimensions. We will discuss the structure of the supersymmetry algebra in more detail in \cite{Komatsu:2024ydh}.

\subsection{Classical supersymmetric vacua}\label{subsec:vacua}
To find the classical vacuum, we set %Lorentz invariant 
$\psi =0$ and look for the minima of the bosonic potential
\begin{equation}
    V_B = \operatorname{Tr} \left[-\frac{1}{4} [X_I,X_J]^2 + \frac{3 \Omega^2}{4^3} X_i X_i + \frac{ \Omega^2}{4^3} X_p X_p + i \frac{\Omega}{3} \epsilon_{ijk} X_i X_j X_k\right].
\end{equation}
The equations of motion (EOM) obtained from varying $X_i$ and $X_p$ are, respectively,
\begin{align}
&\frac{1}{2} [X_I,[X_I,X_i]] + \frac{3 \Omega^2}{4^3} X_i + \frac{i \Omega}{4} \epsilon_{ijk}[X_j,X_k]=0\,,
\\
&\frac12 [X_I,[X_I,X_p]]+\frac{\W^2}{4^3}X_p=0\,,
\end{align}
where $I=1,2,\ldots,10$, $i,j,k=1,2,3$ and $p=4,\ldots,10$. The only solution for $X_p$ is\footnote{See \cite[Sec.3]{Hartnoll:2024csr} for the derivation.}
\al{
X_p = 0\,.
}
As for $X_i$, the solutions can be easily found using the ansatz 
\begin{equation}
    X_p = 0\,,\qquad 
    X_i = \alpha \Omega  J_i\,, \qquad [J_i,J_j] = i \epsilon_{ijk} J_k\,,
\label{eq:saddle ansatz}
\end{equation}
with $J_i$ being the $N\times N$ matrix representations of $SU(2)$ Lie algebra (not necessarily irreducible). In this configuration the bosonic matrices form a 3-dimensional \emph{fuzzy sphere} whose radius has the (dimensionless) length scale $\W$. Plugging the ansatz into the equation of motion we get
\begin{equation}
     \alpha^3 - \frac{1}{2}\alpha^2+ \frac{3}{64} \alpha = 0\,.
\end{equation}
The roots of this polynomial are
% $0,\frac{1}{8},\frac{3}{8}$.
\begin{equation}
     \a=0, \qquad   \a=\frac{1}{8}, \qquad   \a=\frac{3}{8}.
    \label{eq:fuzzy_spheres_saddles}
\end{equation}
%\begin{equation}
 %   \alpha_0 =0, \qquad \alpha_1 = \frac{1}{8}, \qquad \alpha_2 = \frac{3}{8}.
%    \label{eq:fuzzy_spheres_saddles}
%\end{equation}
When $\alpha=%\alpha_0=
0$ we have the trivial vacuum with $X^I=0$ and the on-shell action vanishes. At the non-trivial extrema the potential takes values 
\begin{equation}
\begin{split}
    V_B &= \alpha^2 \Omega^4 \operatorname{Tr}\left[ \frac{\alpha^2}{4} \epsilon_{ijk} \epsilon_{ijl} J_k J_l + \frac{3}{4^3} J_i J_i- \frac{\alpha}{6} \epsilon_{ijk} J_i \epsilon_{jkl} J_l\right] \\
    &=\alpha^2 \Omega^4 \left( \frac{\alpha^2}{2} + \frac{3}{4^3}- \frac{\alpha}{3}\right) \operatorname{Tr}J_i J_i
    \\
    &
    =\begin{cases}
        \frac{5}{3\times 2^{13}}\Omega^4 \op{Tr}J_i J_i \quad (\a=\frac{1}{8})
        \\[5pt]
        -\frac{9}{2^{13}}\Omega^4 \op{Tr}J_i J_i \quad (\a=\frac{3}{8})
    \end{cases}\,.
\end{split}
\end{equation}
The solutions with $\a=\frac38$ are dominant since $\operatorname{Tr}J_i J_i\geq0$. If $J_i$ is the spin-$j$ irrep (and $N=2j+1$), then 
\al{
\operatorname{Tr}J_i J_i = N j(j+1)=N\frac{N^2-1}{4}\,.
}
For a general reducible representation, $J_i$ is block diagonal with $q$ types of blocks, each having multiplicity $n_s$ and size $N_s \times N_s$. Then we have
\begin{equation}
\begin{split}
   \operatorname{Tr}J_i J_i &=\sum_{s=1}^q n_s N_s j_s(j_s+1)=
   \sum_{s=1}^q n_s N_s \frac{N_s^2-1}{4}\,,
   % \\
   % &= N_1 \frac{N_1^2-1}{4}+N_2 \frac{N_2^2-1}{4}+...+N_q \frac{N_q^2-1}{4}\,,
   \qquad N=\sum_{s=1}^q n_s N_s\,.
\end{split}
\end{equation} 
It is straightforward to check that $\operatorname{Tr}J_i J_i$ is maximized by the spin-$\frac{N-1}{2}$ irrep. Thus these solutions are not degenerate and the leading contribution comes from having a single irreducible representation. 
% All together the value of the polarized IKKT action at the saddles is then 
% \begin{equation}
%      S_{IKKT} = - \frac{9}{2^{15}}\Omega^4 \sum_{s=1}^q n_i(N_i^3-N_i).
%      \label{eq:saddle IKKT}
% \end{equation}

Finally we want to see if the saddles preserve supersymmetry. In the background \eqref{eq:saddle ansatz}, the susy transformation of fermions in \eqref{susy_mass_deform} becomes
\begin{equation}
  \delta \psi = \frac{i}{2} (\alpha\W)^2 i \epsilon_{ijk} J_k \Gamma^{ij} \epsilon 
  + \frac{\a 3  \W^2}{8} J_i \Gamma^{123} \Gamma^i \epsilon
  =-\frac{\W^2}{2} \alpha\left(\alpha-\frac{3}{8} \right) i \epsilon_{ijk} J_k \Gamma^{ij} \epsilon\,,
\end{equation}
where we have used
\al{
\Gamma^{1 2 3} \Gamma^k = \frac{1}{2} \epsilon_{i j k} \Gamma^{i j}\,.
}
Therefore, the supersymmetric saddles preserving $\d\y=0$ are $\alpha=0$, the trivial vacuum, and $\alpha = \frac{3}{8}$, corresponding to the minima of the action.

\section{Backreacted geometries}
\label{sec:geometries}
In this section we study Euclidean solutions of type IIB supergravity that are smooth everywhere, have no horizons and have 16 supersymmetries together with $SO(7) \times SO(3)$ symmetry.

%\AV{Should the following piece (before 3.1) be here or in an appendix ?}
Let us review quickly the basics of type IIB supergravity. The bosonic matter content consists of two scalars, the dilaton $\phi$ and the axion $\chi$, and two 3-form field strengths, the Ramond-Ramond (RR) $F_3$ and the Neveu-Schwarz-Neveu-Schwarz (NSNS) $H_3$. 
%There could also be a self dual 5-form flux that we will set to zero in this paper. 
%\SK{slight rewriting.} 
In addition, there is also a self-dual 5-form but we will set it to zero in this paper.
The bosonic part of the action reads
\begin{equation}
    S_{IIB} = \frac{1}{2\kappa^2}\int d^{10}x \sqrt{g}\left(  R -\frac{1}{2} (\partial \phi)^2- \frac{1}{12} e^{-\phi} (H_3)^2- \frac{1}{2} e^{2\phi} (\partial \chi)^2- \frac{1}{12} e^{\phi} (F_3)^2\right).
\end{equation}
%\AV{I am not sure about this paragraph}
This is the Lorentzian action. Since we will construct the Euclidean solution by analytic continuation of a Lorentzian solution, we do not have to use explicitly the Euclidean action. 
Our solution will solve the Lorentzian equations of motion but the metric will have Euclidean signature and the axion and RR 3-form will be purely imaginary. These are the appropriate reality conditions for Euclidean IIB supergravity \cite{Gibbons:1995vg, Ooguri:1998pf}. 
%\XZ{more ref here?}
%In particular we do not have to decide where to put the $i$'s when Wick rotating the matter fields. The solution will take care of that automatically, and some of the forms will be purely real, while some will be purely imaginary.

The field strengths are expressed in terms of 2-forms gauge potentials $B_2$ and $C_2$ as 
\begin{equation}
    H_3 = dB_2, \qquad F_3=d C_2 -\chi H_3.
\end{equation}
There are therefore 3 types of charge in this theory, corresponding to the couplings to the fields $\chi$, $B_2$ and $C_2$. The equations of motion for these fields are 
\begin{equation}
\begin{split}
    d(e^{2\phi}*d\chi)=-e^{\phi} H_3 \wedge *F_3, \qquad
    d(e^{\phi}* F_3)=0, \qquad
    d(e^{-\phi}*H_3)=e^{\phi} d\chi \wedge *F_3.
\end{split}
\end{equation}
Therefore, in addition to the closed 3-forms $dC_2$ and $dB_2$, we can also write a closed 9-form 
% \JP{Maybe we can call this $\tilde{F}_9$.  In \cite{Gibbons:1995vg}, they call $F_9$ the first term.}
\begin{equation}
    \tilde{F}_9 =e^{2\phi}*d \chi + e^\phi B_2 \wedge *F_3,
\end{equation}
%\SK{Isn't it better to write it in terms of $F_3$?} \AV{true, I changed}
and closed 7-forms
\begin{equation}
    \tilde{H}_7 = * \left(e^{-\phi} H_3 - e^\phi \chi F_3\right), \qquad \tilde{F}_7 = * e^\phi F_3.
\end{equation}
The equations of motion then ensure that $d\tilde{F}_9=d\tilde{H}_7=d\tilde{F}_7=0$. This allows us to define conserved ``electric'' charges by integrating those closed forms ($\tilde{F}_9$, $\tilde{H}_7$, $\tilde{F}_7$) on 9-cycles or 7-cycles. Similarly, ``magnetic'' charges are defined by integrating the closed 3-forms ($dC_2$, $dB_2$) on 3-cycles \footnote{Our solution will not have non-contractible 1-cycles and therefore we do not discuss the ``magnetic'' charge of $\chi$, which would measure D7-brane charge.}.
%\AV{Is there a magnetic dual to the axion ?}
%\JP{The dual  $*F_9$ can be integrated on lines therefore it can measure D7-brane charge. I guess the axion is a gauge field for this 1-form field strength $d\chi \sim *F_9$? 
%The axion is not single valued around a D7-brane. Notice that this is OK because the axion is a compact scalar $\chi \sim \chi +2\pi$. However, we do not have 7-branes in our solution because all 1-cycles are contractible.} \AV{footnote added}

Lorentzian type IIB supergravity also has internal $SL(2,\mathbb{R})$ symmetry acting on the scalars and the 3-forms. The latter simply transform as a doublet 
\begin{equation}
    \begin{pmatrix}
        H_3 \\ dC_2
    \end{pmatrix}
    \to \begin{pmatrix}
        d & -c \\ -b & a
    \end{pmatrix} \begin{pmatrix}
        H_3 \\ dC_2
    \end{pmatrix}, \qquad ad-bc=1.
\end{equation}
For the transformation of the axion and dilaton, it is more convenient to group them in the axi-dilaton $\tau = \chi + i e^{-\phi}$ that transforms as 
\begin{equation}
    \tau \to \frac{a \tau + b}{c \tau + d}.
\end{equation}

When we complexify the fields, $SL(2, \mathbb{R})$ gets promoted to $SL(2, \mathbb{C})$. However, in order to obey the reality conditions of Euclidean IIB supergravity, only a subset of this  $SL(2, \mathbb{C})$ is allowed.
%
% In the particular case with one purely real 3-form and one purely imaginary 3-form, We can check that the $SL(2,\mathbb{C})$ transformation with $a=-d=i$ and $b=c=0$ leaves the equations of motion invariant. The net effect is 
% \begin{equation}
%     e^\phi \to -e^\phi, \qquad \chi \to -\chi, \qquad H_3 \to -i H_3, \qquad dC_2 \to i dC_2.
% \end{equation}
% This changes the sign of the axion and dilaton, and exchanges which 3-form is purely imaginary. If we generate a solution with the dilaton being negative, we can use such a transformation to convert it to a regular solution.

% \JP{I think what you mean is that once we allow complex fields then SL(2,R) becomes SL(2,C) and we need to decide what reality conditions we like in Euclidean IIB.}
% \AV{yes}

\subsection{Euclidean solution of IIB supergravity with 16 supercharges}
Lorentzian solutions of IIB supergravity with 16 supercharges and $SO(5,2) \times SO(3)$ symmetry were constructed in \cite{DHoker:2016ujz} and further analysed in \cite{DHoker:2017mds,Corbino:2017tfl,Legramandi:2021uds}. 
%\SK{Added:} 
They are dual to five-dimensional superconformal field theories with the exceptional $F_4$ superconformal symmetry. 

We can obtain the Euclidean version with $SO(7)\times SO(3)$ isometry by performing an analytic continuation as we show in  appendix \ref{app_analytic continuation}. The resulting Euclidean supergravity background has vanishing 5-form field strength and consists of the metric, the axion $\chi$, the dilaton $\phi$, the NSNS 3-form $H_3$ and the RR 3-form $F_3$. It can be written in term of a single function $V(\rho,z)$ that satisfies the equation 
\begin{equation}
    V'' + \Ddot{V} + \frac{2}{\rho} \dot{V} = 0,
    \label{eq:laplace equation}
\end{equation}
where $V' \equiv \partial_z V$ and $\dot{V} \equiv \partial_\rho V$. This is the Laplace equation in a four-dimensional axially symmetric system, where $\rho$ is the radial coordinate and $z$ is the vertical direction. The remaining two angular coordinates are not spacetime coordinates. 

The Einstein frame solution reads\footnote{The Lorentzian solution was also written in terms of a 4d electrostatic potential in \cite{Legramandi:2021uds}. We thank  Nikolay Bobev,Pieter Bomans and Fridrik Freyr Gautason for pointing this to us.}
\begin{equation}
\begin{split}
    ds^2 &= \frac{8}{\mu^{\frac{5}{2}}}\left( \frac{1}{3^3}\frac{ \Delta \dot{V}}{ (-V'') }\right)^{1/4}\left[ \frac{(-V'')}{ \dot{V}} (d\rho^2+dz^2)+3 \rho  d\Omega_6^2 +  \frac{\rho (-V'') \dot{V}}{\Delta} d\Omega_2^2 \right], \\
    e^{\phi} &=  -\mu^3\frac{3 \dot{V}+\rho V'' }{\rho \sqrt{3 \Delta \dot{V}(-V'')}}, \\
     \chi &= -\frac{i}{\mu^3}\frac{3 \dot{V}(V'+\rho \dot{V}')+\rho V' V''}{3 \dot{V} + \rho V''}, \\
     H_3 &= d B_2, \qquad B_2 = -\frac{8}{3\mu} \left(z - \frac{\rho \dot{V} \dot{V}'}{\Delta} \right) \wedge d \Omega_2  \\
    F_3 &= d C_2- \chi H_3, \qquad C_2 = i \frac{8}{3\mu^4}\left(V - \rho \frac{\dot{V}}{\Delta} \left(V' \dot{V}'+3 \dot{V}(-V'') \right) \right)\wedge d \Omega_2, \\
    \Delta &\equiv 3 \dot{V}V''+\rho(\dot{V}'^2+V''^2).
\end{split}
\label{eq:sugra_solution}
\end{equation}
The coordinates $z,\rho$ have dimensions of length, $\mu$ is a mass parameter and $V$ has dimensions of inverse length squared.
This ansatz solves the equations of motion but we still have to impose regularity and positivity of the metric components. This leads to additional constraints on the function $V$. To identify boundaries in the $(\rho,z)$ plane, we look for regions where the $S^2$ and/or the $S^6$ shrink. If we write their respective radii as $R_2$ and $R_6$ we note that %\XZ{changed $\sim$ to $\propto$} 
\begin{equation}
    R_6^3 R_2 \propto \rho^2 \dot{V}.
\end{equation}
Looking back at the metric we see that $\rho\to0$ corresponds to $S^6 \to 0$ and $\dot{V}\to 0$ corresponds to $S^2 \to 0$. To get a smooth solution we also want to impose boundary conditions so that the other components of the metric are finite when one of the spheres shrinks.

We start by looking at the coefficients of the metric when $S^2$ shrinks ($\dot{V} \to 0$ and $\rho \sim 1$). We need $V'' \Delta^{1/3}\to 0$ for the coefficient of $d\rho^2+dz^2$ to be finite, and $V'' \Delta^{-1} \to 0$ for $S^6$ to be finite. The solution is then $V'' \to 0$ and $\Delta \sim 1$.
By using the Laplace's equation we also get  $\ddot{V} \to 0$.

Now we do a similar study in the region where $S^6$ shrinks ($\rho \to 0)$. $S^2$ is finite if $\Delta \sim V'' \dot{V}^{5/3} \rho^{4/3}$, and the 2-dimensional space is finite if $\Delta \sim \frac{\dot{V}^3}{V''^3}$. Taking the ratio to eliminate $\Delta$ we find that $\dot{V} \to 0$, which means, looking at the expression for $\Delta$, that we also have $\Delta \to 0$.

In summary we find that the metric is smooth if in those regions we satisfy
\begin{align}
\begin{split}
    &S^2\ \text{region :} \qquad V'' \to 0, \qquad \Delta \sim 1, \qquad  \dot{V} \to 0, \\
    &S^6\ \text{region :} \qquad \dot{V} \to 0, \qquad \Delta \to 0, \qquad \rho \to 0. 
\end{split}
\end{align}
The $S^6$ region is just the $z$ axis. For the $S^2$ region we need more information about the implicit curves defined by $\dot{V}=0$. We can compute the slope of those curves as 
\begin{equation}
    m = \frac{\partial_z \dot{V}}{\partial_\rho \dot{V}}= \frac{\dot{V}'}{\Ddot{V}}.
\end{equation}
The denominator vanishes due to the the Laplace equation together with $\dot{V}\to 0$ and $V''\to 0$. However the numerator is finite due to $\Delta \sim 1$. Therefore the slope in the $(\rho,z)$ plane is zero and those curves are at constant $z$.
This allows us to use an electrostatic analogy where $\partial_\rho V=0$ is interpreted as the vanishing of the tangential component of the electric field on the boundary of a conductor. To respect the axial symmetry of the system, those conductors need to be 3 dimensional balls that are infinitely thin in the $z$ direction. They are the higher dimensional analog of the 2d conducting disks from \cite{Lin:2005nh} in a 3d axially symmetric system. In the following we will also refer to the balls as ``disks''.

Since the spheres go to zero at different points in the $(\rho,z)$ plane, we can construct cycles by connecting such points with a curve that passes through a region where the spheres are finite. In particular we can draw a segment between two disks over which we tensor the $S^2$, and that defines a 3-cycle $\Sigma_3$\footnote{This is similar to a sphere, which is a 2-cycle, that can be defined by tensoring an $S^1$ over a segment, with the $S^1$ shrinking at the endpoints, that are then the north and south poles of the sphere.}. Similarly we draw a curve around a disk with endpoints on the $z$ axis, and, tensoring the $S^6$, that defines a 7-cycle $\Sigma_7$. This construction is depicted on Figure \ref{fig:disks_plane}.

This is not enough to guarantee that the metric and dilaton are positive everywhere. We also need boundary conditions at infinity that are set by including a background potential that grows at infinity. The simplest possibility is to choose a harmonic polynomial. As we show in the next section, the polynomial that is relevant for the polarized IKKT matrix model is 
\begin{equation}
\label{eq: Vbg}
    V_{bg}(\rho,z) = -\eta z \mu^3  +\frac{\mu^5}{2^7}(
z \rho^2-z^3)\,,
\end{equation}
where the term proportional to the dimensionless parameter $\eta$ has no effect on the metric but can affect the different matter fields. We will later see that $\eta$ is related to a constant term that can be added to the action of the matrix model.
%\SK{Added.}
The different positivity conditions are satisfied in the physical region $\rho \geq 0$ and $z \geq 0$, as we show in appendix \ref{app_regularity}.
%\AV{added :}
Since the $S^6$ shrinks when $\rho=0$, the geometry terminates smoothly there and there is no physical boundary. Similarly we would like the geometry to stop at $z=0$ smoothly, which can be done by introducing a grounded plane (or infinite disk) at $z=0$, making the $S^2$ shrink there.

% Still following \cite{Lin:2005nh}, we assume that there is a grounded plane (or infinite disk) at $z=0$. \AV{Can we say here that the existence of that plane is required for the geometry to terminate smoothly at $z=0$ ? Positivity conditions would be violated for $z<0$. } %\AM{Why ?} \AV{among all the supergravity solutions that have the correct symmetry, only a subset with specific boundary conditions are expected to be dual to the matrix model. Since this is very close to BMN we try to choose similar boundary conditions, and we see that we get a backreaction in a D instanton background so it's consistent. More conservatively, with this choice of background potential the region $z<0$ is unphysical because the regularity conditions for the metric and dilaton are not obeyed. And a different choice of background potential would have nothing to do with the D instanton. }

This concludes the presentation of the supergravity solution. It is determined by choosing a configuration of conducting disks, of respective charges $Q_s$, centered on the $z$ axis at positions $z_s$. Note that the radii of the disks are not free parameters for the following reason: We want the metric to be smooth, and therefore the electric field should not diverge at the tip. This requires the charge density to vanish at the tip, which is a constraint that is possible to satisfy in a background potential. Physically, assuming $Q_s\geq0 $, the electric field produced by the charge of the disk, pushing outward, and the electric field from the background potential, pushing inward, need to cancel at the tip of the disks. This allows to solve for the radii $R_s$ in terms of the charges $Q_s$ and positions $z_s$, which are then the only  free parameters of the solution. 

Once the configuration of disks is chosen, one can solve the four-dimensional Laplace equation to get the function $V(\rho,z)$, which in turns gives the full supergravity solution.

\subsection{Asymptotic region}
\label{sec:asymptotics}
%\AV{I did some rewriting here to consider the first multipole moments.}
Here we study the solution in the asymptotic region $r\to \infty$, where $r^2 \equiv z^2 + \rho^2$. We use a multipole approximation to solve the electrostatic problem in that limit. Using the method of images we find that the leading contribution from the disks is a dipole. Together with the background potential this gives 
\begin{equation}
\label{eq:asymptotic_expansion}
V(\rho,z) = -\eta  z \mu^3  +\frac{\mu^5}{2^7}(
z \rho^2-z^3)
+\frac{1}{2 \pi^2} \frac{z P}{(z^2+\rho^2)^2}+ \frac{1}{4 \pi^2} \sum_{l=1}^\infty q_{2l+1} \frac{U_{2l+1}\left( \operatorname{cos}\theta \right)}{(z^2+\rho^2)^{l+3/2}},
\end{equation}
where $P = 2\sum_{s=1}^q z_s Q_s$ is the dipole moment for a configuration of $q$ conducting disks of charges $Q_s$ and distances $z_s$ from the grounded plane. The $q_l$ for $l \geq 3$ are the higher multipole moments, $U_l$ are the Chebyshev polynomials of the second kind and $\operatorname{cos}\theta = \frac{z}{\sqrt{z^2+\rho^2}}$. We only have odd terms in $l$ because the electrostatic system is antisymmetric under $z \to -z$. Asymptotically, the solution reads
% \JP{Why are you expanding in $P$? Shouldn't we expand in $1/r$?} \AV{I did both and got the same result, but I agree the expansion in $1/r$ is more clear.
\begin{equation}
    \begin{split}
        ds^2 &= dz^2+d\rho^2 + z^2 d\Omega_2^2 + \rho^2 d\Omega_6^2+\mathcal{O}(r^{-4}), \\
        e^\phi &= \frac{2^{14} P}{\pi^2 \mu^7 r^8}+\mathcal{O}(r^{-10}),\\
        \chi &= i \frac{\pi^2 \mu^7 r^8}{2^{14} P}+ \mathcal{O}(r^{6})\\
          dC_2 &=  -i\mu\  z^2 dz \wedge d\Omega_2+\mathcal{O}(r^{-6}), \qquad H_3 = \mathcal{O}(r^{-6})
    \end{split}
    \label{eq:D-Instanton background}
\end{equation}
Remarkably, the space is asymptotically flat. This is  the solution \cite{Gibbons:1995vg,Ooguri:1998pf} dual to the pure IKKT matrix model with the addition of $dC_2$ flux. 
%\AV{added comment about the flux being constant}
This flux is in fact constant because we can use $(x^1,x^2,x^3)$ as the Cartesian coordinates associated to the $S^2$ with radius $z$. Then 
\begin{equation}
    dC_2 = -i\mu \ dx^1 \wedge dx^2 \wedge dx^3 + \mathcal{O}(r^{-6}).
\end{equation}
%(without the mass deformation). \JP{Do you agree?} \AV{yes, but with particular values for $\phi_\infty$ and $\chi_\infty$ that are free parameters in \cite{Gibbons:1995vg}. This is true because all the term $\mathcal{O}(r^\#)$ comes with positive powers of $\mu$.}

Let us make a few comments on the asymptotic form of the solution \eqref{eq:D-Instanton background}. First, setting the dipole moment $P$ to zero makes the axion diverge. However, this does not pose a problem to our analysis since, as we will discuss in section \ref{subsec:quantization}, the dipole moment is identified with the size of the matrices which is always nonzero.
Second, note that asymptotically we have $i \chi = -e^{-\phi}$. This relation is still obeyed by the first subleading terms in the asymptotic expansion, proportional to $q_3, q_5$ and $q_7$. We can measure the first deviation as
\begin{equation}
\label{eq:deviation_chi_phi}
    e^{-\phi} +i \chi  =  \frac{3z^2 + \rho^2}{64}\mu^2- \eta + \mathcal{O}(r^{-2}),
\end{equation}
where we recognize the bosonic mass term of the polarized IKKT model \eqref{mass-deformed-IKKT}.

\paragraph{S-duality.} Our solutions are in a frame where we only have RR flux and no NSNS flux in the asymptotic region. We can go to another frame by using $SL(2,\mathbb{C})$ transformations. In particular 
%\AV{Here I precised that this particular transformation is in fact S-duality} 
the S-duality transformation with $a=d=0$ and $c=b=i$ will map $H_3 \to i dC_2$, and $dC_2 \to i H_3$, preserving the reality conditions for Euclidean IIB supergravity. In this new frame, the NSNS flux is purely real, and the RR flux is purely imaginary. 
%\SK{Below, I made several changes. Please check if I'm not saying something incorrect.} \AV{everything is correct}
Setting $\eta = \frac{1}{2}$, we get the solution of the form
% \AV{I use the notation "$\simeq$ to mean "up to terms subleading in $r$, is it clear ?}
% \JP{Not really. Can you write $+O(r^{-\#})$? }
\begin{equation}
    e^\phi =  1-\frac{\mu^2}{32}(3z^2 + \rho^2) + \mathcal{O}(r^{-2}) , \qquad \chi = -i e^{-\phi}+\mathcal{O}(r^{-8}), \qquad H_3 = -\mu dx^1\wedge dx^2 \wedge dx^3+\mathcal{O}(r^{-6}).
    \label{eq:cavity}
\end{equation}
In this frame, we can set all the multipoles to zero without encountering divergences. Doing so kills the subleading terms $\mathcal{O} (r^{-\#})$ in \eqref{eq:cavity} and makes the solution identical to the recently found ``cavity'' solution \cite{Hartnoll:2024csr}. 
%\AV{comment added, I'm not sure if it brings anything interesting to the discussion} 
 On the other hand, in our original frame \eqref{eq:D-Instanton background}, setting multipoles to zero would make the axion diverge\footnote{The axion diverges also in the cavity solution \eqref{eq:cavity} but only at the boundary of the ellipsoid $3z^2+\rho^2 \leq \frac{32}{\mu^2}$.} everywhere. 
 %\AV{Here I rephrased a bit} 
 We will comment more on the comparison between the two frames in section \ref{sec: probe D1}. Note that taking the asymptotic limit does not commute with S-duality. One should first take the $SL(2,\mathbb{C})$ transformation in \eqref{eq:sugra_solution} and then expand the solution asymptotically.

% \AV{Since asymptotically we have the same metric and dilaton as the $\mu=0$ solution, does the string frame metric still describe a wormhole ? Shall we comment on that ?}
% \SK{I thought a bit about it, but I don't think we can say anything since Ooguri-Skenderis solution is actually not a wormhole. Basically if we take a near horizon limit, we lose one of the two asymptotic boundaries.}
% \AV{I see, we have the solution 4.2 in OS rather than 3.7, loosing the "+1" in the dilaton the string metric does no longer have the $r \to \frac{1}{r}$ isometry.}

\subsection{12D uplift}
In the analogous case of the BMN matrix model, the asymptotic region is a 11 dimensional pp-wave, which is not manifest here. However Euclidean IIB supergravity can also be formulated in 12D \cite{Tseytlin:1996ne} using 
\begin{equation}
\label{eq:12D_uplift}
    ds_{12}^2 = ds_{10}^2 + M_{ij}dy^i dy^j, \qquad M \equiv e^\phi\begin{pmatrix} \chi^2 + e^{-2\phi}& -\chi \\ -\chi & 1 \end{pmatrix},
\end{equation}
%\AV{added:}
where $ds^2_{10}$ is the 10 dimensional metric of the IIB solution and $(y^1,y^2)$ are coordinates on a non-dynamical 2-torus.
Using our asymptotic solution \eqref{eq:D-Instanton background}
%\AV{footnote added}
% \AV{It turns out that to get this you need to expand $e^\phi$ and $\chi$ up to terms that contain $P^2$. This comes late in the $1/r$ expansion because first you encounter the higher multipoles $q_3$, $q_5$ $q_7$. All those terms cancel when we do the uplift. This is because the first term in the expansion of $e^\phi- \frac{i}{\chi}$ is at $P^2$. I'm not sure what is the best way to explain that, should we write all the terms up to $P^2$ ?} 
we get\footnote{The coefficient of $dy_1^2$ comes from the first non zero term in $\chi^2+e^{-2\phi}$, measured by \eqref{eq:deviation_chi_phi}. Due to high degree of cancellations we need to keep many terms in the asymptotic expansion. The final result is obtained by expanding $e^{\phi}$ up to terms of order $r^{-16}$ and $\chi$ up to terms of order 1.}
\begin{equation}
\label{eq:12D_metric_P}
    ds_{12}^2 =ds_{10}^2 - 2i dy_1 dy_2 + \left(\frac{\mu^2}{32} (3z^2+\rho^2)-2\eta \right) dy_1^2 + \frac{2^{14} P}{\pi^2  \mu^7 r^8}dy_2^2 + \mathcal{O}(r^{-2}).
\end{equation}
The $dy_2^2$ term that we displayed is the leading contribution for this metric component.
%\AV{If we remove the discussion about DLCQ shall we also remove this comment ?}
%we will discuss below why we chose to show it.
Identifying a time component $t = -i y_1$ all the components of the metric are real. Let us now discuss the matter fields.
A 10d solution with vanishing 5-form can also be obtained from a 12d solution with only a 4-form $F_4$ such that 
\begin{equation}
    (F_4)_{y_1 \mu \nu \rho}= ( dC_2)_{\mu \nu \rho}, \qquad (F_4)_{y_2 \mu \nu \rho}= (H_3)_{\mu \nu \rho}.
\end{equation}
We then get 
\begin{equation}
    F_4 =  -\mu z^2 idy_1 \wedge dz \wedge d\Omega_2.
\end{equation}
Hence, writing $y \equiv y_2$, we find that the asymptotic form of our solution is the 12 dimensional pp-wave background 
\begin{equation}
\label{eq:12D_metric}
\begin{split}
    ds_{12}^2 &= 2dt dy + \sum_{i=1}^3dx^i dx^i + \sum_{p=4}^{10}dx^p dx^p - 2dt^2 \mu^2 \left( \frac{3}{4^3} x^i x^i + \frac{1}{4^3}x^p x^p -\eta \right), \\
    F_4 & = \mu dt \wedge dx^1 \wedge dx^2 \wedge dx^3,
\end{split}
\end{equation}
where $z^2 = \sum_{i=1}^3 x^i x^i$, and $\rho^2 = \sum_{p=4}^{10} x^p x^p$.
We checked that this indeed satisfies the 12D Einstein equations\footnote{The normalization for $p$ forms is \begin{equation}
    S =  \frac{1}{16 \pi G} \int d^dx \sqrt{g}\left( R - \frac{1}{2(p+1)!} F^{\mu_1...\mu_{p+1}}F_{\mu_1...\mu_{p+1}} \right),
\end{equation}
giving the stress energy tensor 
\begin{equation}
    T_{\mu \nu} \equiv-\frac{16 \pi G }{\sqrt{g}} \frac{\partial S_m}{\partial g^{\mu \nu}}= \frac{1}{12}\left( F_{\mu abc} F_\nu^{\ abc} - \frac{1}{8}g_{\mu \nu}F_{abcd}F^{abcd} \right).
\end{equation}
In these conventions the field equations are $ R_{\mu \nu} -\frac{1}{2}R g_{\mu \nu} = T_{\mu \nu}$.}.

% \AV{not sure about the following paragraph}
% Note that when going from \eqref{eq:12D_metric_P} to \eqref{eq:12D_metric} we dropped the $P$ dependent term in $dy^2$. This is in fact not harmless since $t$ and $y$ are then interpreted as lightcone coordinates, which makes the compactification singular. See \cite{Hellerman:1997yu} for a similar discussion in the case of the 11D plane wave. Usually this can be regularized by considering a slightly spacelike compactification, which can be done by adding a term $\sim \epsilon dy^2$ in the metric. In our case we note that the leading dependence in $P$ gives such a term and automatically regularizes the lightlike compactification.

We can also write the 12D uplift in general and, using \eqref{eq:sugra_solution} and \eqref{eq:12D_uplift}, we get
\begin{equation}
\begin{split}
    ds_{12}^2 &= \left(\frac{1}{3 \rho^2 \Delta \dot{V} (-V'')} \right)^{1/2} \left[ -\mu^3(3 \dot{V} + \rho V'') dy^2 -2 \tilde{\Delta} dt dy - \frac{1}{\mu^3}\frac{\tilde{\Delta}-3\rho^2 \dot{V}(-V'')\Delta}{3 \dot{V}+\rho V''} dt^2\right],\\
    &+ \frac{8}{\mu^{\frac{5}{2}}}\left( \frac{1}{3^3}\frac{ \Delta \dot{V}}{ (-V'') }\right)^{1/4}\left[ \frac{(-V'')}{ \dot{V}} (d\rho^2+dz^2)+3 \rho  d\Omega_6^2 +  \frac{\rho (-V'') \dot{V}}{\Delta} d\Omega_2^2 \right] \\
    F_4 &= -i dC_2 \wedge dt + dB_2 \wedge dy,\\
     B_2 &= -\frac{8}{3 \mu} \left(z - \frac{\rho \dot{V} \dot{V}'}{\Delta} \right) \wedge d \Omega_2 , \qquad C_2 = i \frac{8}{3 \mu^4}\left(V - \rho \frac{\dot{V}}{\Delta} \left(V' \dot{V}'+3 \dot{V}(-V'') \right) \right)\wedge d \Omega_2, \\
    \tilde{\Delta} &= 3 \dot{V} (V'+\rho \dot{V}')+\rho V' V'', \qquad \Delta = 3 \dot{V}V''+\rho(\dot{V}'^2+V''^2).
\end{split}
\end{equation}

\subsection{Quantization of fluxes}\label{subsec:quantization}
%We found geometries parametrized by continuous parameters, the charges of the disks $Q_i$ and their positions $z_i$. However, to make contact with $N \times N$ matrices, we need a discrete set %\AV{discretized instead of finite ?} a finite number 
%of solutions. \AV{need some rephrasing} The puzzle is resolved by imposing charge quantization on the fluxes.
We found geometries parametrized by continuous parameters, the charges of the disks $Q_s$ and their positions $z_s$. On the other hand, as we saw in section \ref{subsec:vacua}, the polarised IKKT model has a discrete set of classical vacua. 
%\SK{changed parameters to vacua} 
The way to connect the two descriptions is to impose the Dirac charge quantization condition on the fluxes present in the geometry \cite{Lin:2004nb}. We will see that in practice this will quantize the charges of the disks, as well as their respective distances, their respective difference of potentials and the leading dipole moment. At the end the classical supergravity approximation is valid when those quantum numbers are large. Hence those quantization conditions effectively become invisible, as we will see explicitly below. Nevertheless we learn how those charges
%\SK{changed to charges} 
are mapped to the representations of $SU(2)$ characterizing the classical vacua \eqref{eq:saddle ansatz} corresponding to saddle points of the matrix integral.

%\AV{I proposed this new version : We found geometries parametrized by continuous parameters, the charges of the disks $Q_i$ and their positions $z_i$. However $N \times N$ matrices have a discrete set of parameters. The way to connect the two descriptions is to impose the Dirac charge quantization condition on the fluxes present in the geometry \cite{Lin:2004nb}. We will see that in practice this will quantize the charges of the disks, as well as their respective distances, their respective difference of potentials and the leading dipole moment. At the end the classical supergravity approximation is valid when those quantum numbers are large, hence those quantization conditions become invisible, as we will see more explicitly below. However, along the way, we learn how those quantum numbers are mapped to the SU(2) matrices \eqref{eq:saddle ansatz} corresponding to saddle points of the matrix integral.}
 
To quantize the fluxes we study the  equations of motion and Bianchi identities to identify conserved charges, as shown at the beginning of this section. We have closed $p$-forms with $p=3,7,9$. As we explained, the geometries have 3-cycles and 7-cycles. We can then compute the fluxes of closed 3-forms on $\Sigma_3$ and of closed 7-forms on $\Sigma_7$. The flux of the closed 9-form can be computed on a large $S^9$ in the flat space asymptotic region. 
We start with the latter that computes the instanton charge.
We recall 
\begin{equation}
    \tilde{F}_9 =e^{2\phi}*d \chi + e^\phi B_2 \wedge *F_3
\end{equation}
%\AV{added : }
Much like other fundamental $Dp$-branes, the fundamental instanton has unit charge. Thus, the integral of $\tilde{F}_9$ counts the number of instantons. Evaluating the integral on $S^{9}$ at infinity, we find\footnote{The integral on $S^{9}$ of radius $R$ gives $\frac{1}{(2\pi)^8 \alpha'^4 g_s}\int_{S^9} \tilde{F}_9 = \frac{2^7}{3\pi^5 \mu^7 \alpha'^4 g_s}  P+ \mathcal{O}(R^{-6})$. Sending $R\to\infty$ gives \eqref{eq:N-1P}.}
\begin{equation}
    N_{(-1)} = \frac{1}{(2\pi)^8 \alpha'^4 g_s}\int_{S^9} \tilde{F}_9 = \frac{2^7}{3\pi^5 \mu^7 \alpha'^4 g_s}  P.
    \label{eq:N-1P}
\end{equation}
%\SK{removed $\mathcal{O}(R^{-6})$ and moved it to footnote.} 
% \JP{The error decays with the radius? Can we write an equality using $\lim_{R_{S^9} \to \infty}$?} \AV{With the next term I get $N_{(-1)} \sim P\left(1+\frac{P}{R^6 \mu^4}+...\right)$. The octopole $q_3$ gives a first contribution $\sim \frac{q_3 P}{\mu^4 R^8}$ an is also suppressed.}
The dipole moment is therefore quantized. Note that this value of $N_{(-1)}$ can also be checked by matching the coefficient of the dilaton in the D-instanton solution \cite{Itzhaki_1998}. 
%\AV{changed reference, I matched with $p=-1$ in IMSY rather than Ooguri-Skenderis since the latter does not provide the numerical factor of $e^\phi = \#/r^8$, while the former does} 

%We also find agreement.

We now compute the other fluxes. The closed 7-forms are
\begin{equation}
    \tilde{H}_7 = * \left(e^{-\phi} H_3 - e^\phi \chi F_3\right), \qquad \tilde{F}_7 = * e^\phi F_3,
\end{equation}
and the closed 3-forms are $H_3$ and $dC_2$. %\AV{added / rephrased:} 
Integrating those fluxes counts the number of $D1$ and $D5$ branes, as well as the number of $F1$ fundamental strings and of $NS5$ fivebranes, as we see from the quantization conditions  
\begin{equation}
\begin{split}
   \frac{1}{(2\pi)^2 \alpha'}\int_{\Sigma_3} H_3 &= N_{NS5}, \qquad \frac{1}{(2\pi)^6 \alpha'^3 g_s}\int_{\Sigma_7} \tilde{F}_7 = N_{D_1}, \\
   \frac{1}{(2\pi)^2 \alpha' g_s}\int_{\Sigma_3} dC_2 &= N_{D5}, \qquad \frac{1}{(2\pi)^6 \alpha'^3}\int_{\Sigma_7} \tilde{H}_7 = N_{F_1},
\end{split}
\end{equation}
where the labels of the integers $N_{\rm label}$ refer to the type of brane carrying the charges.
The integral on $\Sigma_3$ can be done close to the axis at $\rho=0$.
%$S^6$ region when $\rho \to 0$. 
There we expand
\begin{equation}
    %S^6\ \text{region :} 
    \rho\to 0 :  \qquad V(\rho,z) = f(z) - \frac{f''(z)}{6} \rho^2+ \frac{1}{120} f^{(4)}(z)\rho^4+...,
    \label{eq:V close to axis}
\end{equation}
where the terms in the Taylor expansion are set by using Laplace's equation.
If we integrate $H_3$  we get \begin{equation}
     4\pi^2 \alpha' N_{NS5} = \int_{\Sigma_3} i H_3 =  \frac{8}{3\mu} \int_{z_{s-1}}^{z_{s}} d\Omega_2 dz = \Omega_2 \frac{8}{3\mu} d_s, 
\end{equation}
where $d_s \equiv z_{s}-z_{s-1}$ is the distance between the disks, that is therefore quantized.
To do the integrals on the 7-cycle $\Sigma_7$ we can pick a contour close to the disk and expand 
\begin{equation}
  %  S^2\ \text{region :}
    z\to z_s : \qquad V(\rho,z) = V_s + g(\rho)(z-z_s)-\frac{2\dot{g}(\rho)+\rho \ddot{g}(\rho)}{6 \rho}(z-z_s)^3+...
    \label{eq:V close to disks}
\end{equation}
Integrating $\tilde{F}_7$ we get  
\begin{equation}
  (2\pi)^6 \alpha'^3 g_s N_{D1} = \int_{\Sigma_7} i \tilde{F}_7 = \frac{2560}{\mu^6} \int_0^{\rho_s} d\Omega_6  d\rho\ \rho^3 \partial_\rho \left(  \partial_z V(z^+)-\partial_z V(z^-) \right) = \frac{2560}{\mu^6} \Omega_6 \frac{Q_s}{\Omega_2},
\end{equation}
and therefore the charges of the disks are quantized. Here by $z^+$ we mean $z+ \epsilon$ with $\epsilon>0$ and similarly $z^- \equiv z-\epsilon$.
To summarize, we get the following quantization conditions \footnote{The $``s``$ in $g_s$ is not a label for the disks, but the standard name for the string coupling constant.} 
%\AV{using $s$ for the disks labels here is confusing because we also have the string coupling $g_s$. I added a footnote}
\begin{equation}
    \begin{split}
    N_{(-1)} = \frac{2^7}{3\pi^5 \mu^7 \alpha'^4 g_s}  P\,, \qquad 
        N_{NS5,s}=  \frac{8 }{3 \pi \mu \alpha'}  d_s\,, \qquad N_{D1,s} =  \frac{2^5 }{\pi^4 \mu^6\alpha'^3 g_s}Q_s\,.
    \end{split}
    \label{eq:electrostatic parameter matching}
\end{equation}
Now using $P=2 \sum_s Q_s \sum_{t\leq s}d_t$, we get that all numerical factors cancel and 
\begin{equation}
    N_{(-1)} =  \sum_s N_{D1,s} \sum_{t\leq s}N_{NS5,t},
\end{equation}
which expresses the partitions of $N_{(-1)}$ in the integers $\sum_{t\leq s}N_{NS5,t}$ . We can compare this with expectations from the polarized IKKT matrix model. The vacua are fuzzy spheres with 3 matrices in the adjoint of $SU(2)$. Such a matrix can generically be written as
% \begin{equation}
%     J^i = \bigoplus_{n=1}^{N_0} n_i J^i_{(N_i)},
% \end{equation}
\begin{equation}
    J^i = \bigoplus_{s=1}^{q} \mathbb{1}_{n_s} \otimes J^i_{(N_s)}\,,
    \qquad
    i=1,2,3\,,
    \qquad
    N=\sum_{s=1}^q n_s N_s\,,
\label{eq:fuzzy_spheres_saddles_in_J^k}
\end{equation}
where $J^i_{(N_s)}$ is the $N_s=2j_s+1$ dimensional (spin-$j_s$) irrep, and $n_s$ are the multiplicities of those irreps. There are $P(N)$ such configurations. It is then natural to identify 
\al{
N \leftrightarrow N_{(-1)}\,,
\qquad
n_s \leftrightarrow N_{D1,s}\,,
\qquad 
N_s \leftrightarrow \sum_{t\leq s}N_{NS5,t}\,. 
\label{eq:parameter identification}
}
Therefore the gravity solution dual to this vacuum corresponds to a configuration with $q$ disks at positions $z_s= \sum_{t\leq s}d_t$ with respective charge $Q_s$, see Figure \ref{fig:vacua}.

Let us now discuss the two quantization conditions that we omitted. The $\tilde{H}_7$ integral yields
\begin{equation}
\label{eq:F1zero}
  (2\pi)^6 \alpha'^3 N_{F1} = \int_{\Sigma_7} \tilde{H}_7 = \frac{2^9}{\mu^9}\Omega_6  \int d\rho\ \rho^3 \left(3\rho (\partial_z \partial_\rho V)^2- \partial_\rho ( \partial_z V)^2 \right) \big|^{z=z_s+\epsilon}_{z=z_s-\epsilon} = 0,
\end{equation}
and vanishes because the integrand takes identical values above and below the disk. On the other hand the integral of $dC_2$ gives
\begin{equation}
\label{eq:ND5}
  4\pi^2 \alpha' g_s N_{D5} =  \int_{\Sigma_3} dC_2 = \frac{8}{3\mu^4}  \int_{z_{s-1}}^{z_{s}} d\Omega_2 dz \partial_z V(z,0)  = \Omega_2 \frac{8}{3\mu^4} ( V_s - V_{s-1}).
\end{equation}
where $V_s$ gives the (constant) potential of the disk $s$. This shows that the difference of potentials between neighboring disks must be quantized in units  proportional to $g_s N_{D5}$. This raises an apparent puzzle since  the electrostatic problem is already fully determined and it is unclear that we could impose this additional constraint. A resolution to this puzzle comes from the fact that the supergravity approximation is valid when $g_s \to 0$ and the quantum number $N_{D5}$ is large. Thus we can choose the product $g_s N_{D5}$ so that $V_s-V_{s-1}$ is arbitrarily close to any real number, which is determined by solving the electrostatic problem. Note that the same comments apply also to quantizations imposed by $N_{-1}$ and $N_{D1}$, both of which come with a factor of $g_s$: strictly speaking, quantizations of these fluxes do not constrain the parameter space of the solutions in the supergravity limit.
%\SK{Changed}

%\AV{add ? : As advertised at the beginning of this section, this is how this quantization condition becomes invisible in the classical regime.}

%\XZ{add this?}The attentive readers may realize that the argument for continuous $V_i-V_{i-1}$ through small $g_s$ also applies to $N_{(-1)}$ and $N_{D1}$.... In other words, there is no consistent quantization for finite $N_{\rm label}$'s, and only at large $N$'s can the electrostatic problem be consistently defined. The outcome of this quantization and ``de-quantization'' (through large $N$) is that we find the relations among charges and the electrostatic parameters, encapsulated by \eqref{eq:electrostatic parameter matching}.

%\JP{I am confused. Why don't we say the same words about $P \propto g_s N_{-1}$ or $Q \propto g_s N_{D1}$?}
%\JP{Do you mean that SUGRA corresponds to $N_{D5} \to \infty$ with fixed $g_s N_{D5}$? } \AV{Is it reasonable ? I don't know what $N_{D5}$ could correspond to in the matrix model... }

% Note that the prefactor is identical to the prefactor of $\int H_3$, apart from the factor of $g_s$.  Therefore we have the constraint 
% \AV{I don't know if that comment about the identical prefactor is useful because changing this relation with a number would not change our discussion}
% \begin{equation}
% \label{eq:ND5}
%    \Delta V_i = g_s \frac{N_{D5}}{N_{NS5}}d_i 
% \end{equation}

\subsection{Validity of the supergravity approximation}
To trust the supergravity approximation we need
small curvature of the string frame metric $ds^2_{(s)} \equiv e^{\phi/2} ds^2$ to avoid stringy higher curvature corrections,
%to the low energy Einstein-Hilbert effective action, 
and  small string coupling $g_s e^\phi$  to avoid string loop corrections. To probe the curvature we use the Ricci scalar $R$ of the string metric in string units, i.e. we study $\alpha' R$.

Let us first study those quantities in the asymptotic region where we use the expansion \eqref{eq:asymptotic_expansion}. We get 
\begin{equation}
    r\to \infty :  \qquad R \alpha' \sim \frac{r^2}{\sqrt{g_s N} \alpha'} \sim \frac{r^2}{\alpha'^2 \mu^2\sqrt{ \lambda}}, \qquad g_s e^\phi \sim  \frac{g_s^2 N \alpha'^4}{r^8} \sim \frac{\mu^8 \alpha'^8 \lambda^2}{N r^8},
    \label{curvature dilaton infinity}
\end{equation}
where $\lambda \equiv N g_{YM}^2/\mu^4$ is the (dimensionless) 't Hooft coupling. The dilaton decays at infinity but is finite close to the axis and on the disks. We will estimate the value of the dilaton in those regions. We also estimate the curvature since the disks should be inside the trustworthy region. Since we trust the approximation \eqref{curvature dilaton infinity} from infinity down to
%\XZ{how about "from infinity down to"} 
$r \sim \operatorname{max}(\rho_s,z_s)$, we also need to check that the conditions coming from the approximation near the disks are not weaker than the condition coming from the asymptotics at that point. 
\begin{figure}[h]
    \centering
\includegraphics[scale=0.25]{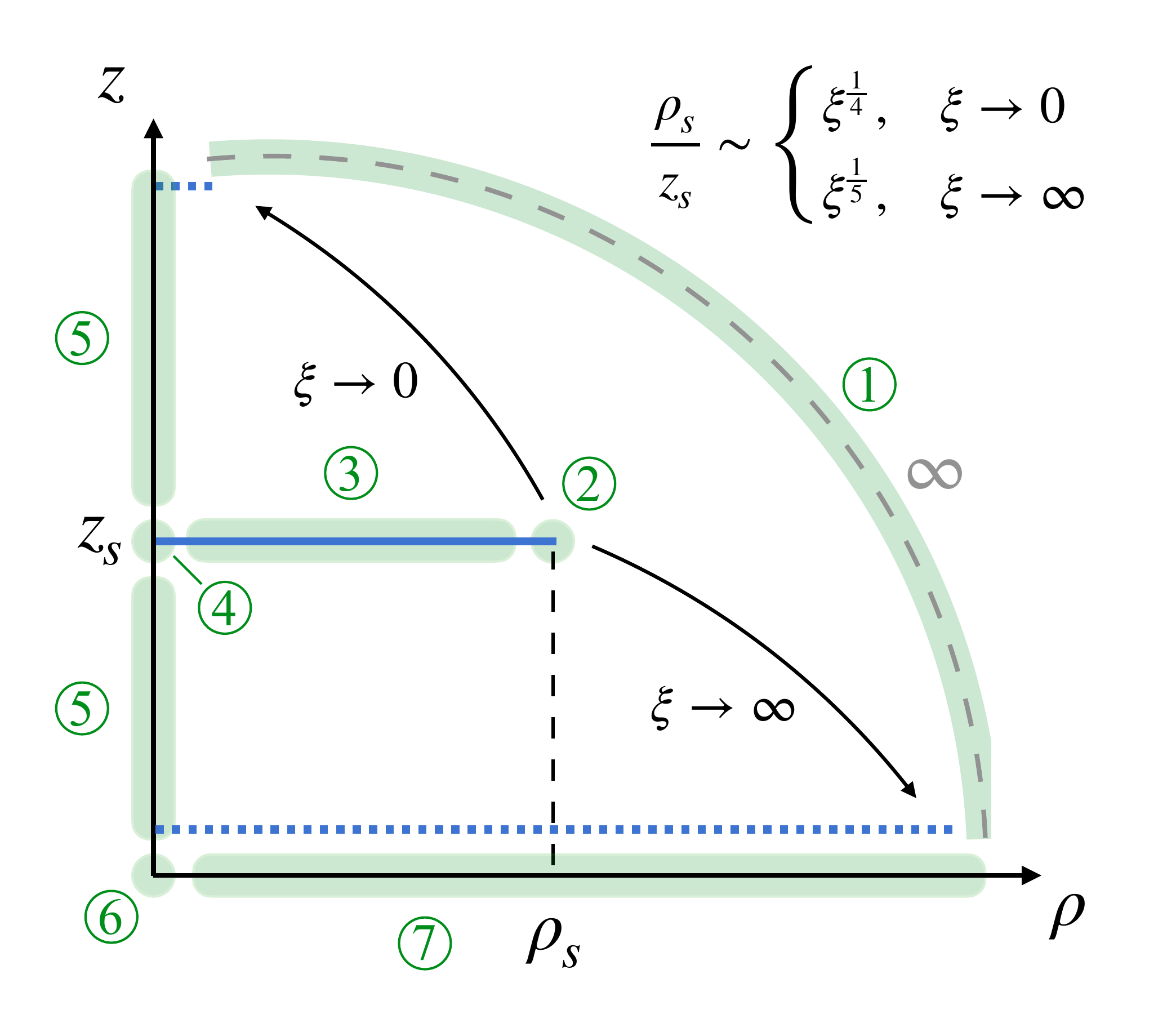}
    \caption{Summary of our analysis of the supergravity regime. We study the scaling of the string coupling $g_s e^\phi$ and the curvature probed by the Ricci scalar in regions \circled{1} to \circled{7}. The scaling behaviors are given in table \ref{tab:summary of validity analysis}. We find that the strongest conditions come from infinity and the tip of the disk (region \circled{1} and \circled{2}).}
    \label{fig:sugra_regime}
\end{figure}

%\AM{So in the case where the condition \eqref{curvature dilaton infinity} at $r \sim \mathrm{max}(\rho,z_s)$ is stronger than the approximation near the disks \eqref{R and phi}, this means that we shouldn't trust the approximation near the disks?} \AV{The approximation is fine, but the leading scaling will come from the stronger condition, with the gluing interpolating between the two, i.e. we want to understand if the curvature and dilaton in the gluing region is larger than at the tip. If they are larger in the asymptotic region evaluated as $\operatorname{max}(\rho_s,z_s)$ we conclude that the stronger constraint will come from that.}

At finite distance, we study the case of a single disk (and its image) and write the potential
\begin{equation}
    V(\rho,z) = V_{bg}(\rho,z) + \frac{1}{4 \pi^2 \rho} \int_0^{\rho_s} dr f_s(r)  \left(\frac{\rho-r}{(\rho-r)^2+(z-z_s)^2}-\frac{\rho-r}{(\rho-r)^2+(z+z_s)^2}  \right),
\end{equation}
where $f'_s(r) = -2 \pi r \sigma_s(r)$ with $\sigma_s(r)$ the charge density of the $s$ disk. This expression for $V(\rho,z)$ can be obtained from $\eqref{potential V}$ by integrating by parts. Then, we plug this expression in the equations for the dilaton and the curvature and expand around the regions of interest. 
%\XZ{added:} 
The main input to specify here is the function $f_s(r)$. Since only the integral of $f_s(r)$ is needed, we assume an even distribution $f_s(r) \sim Q_s / \r_s$ and this suffices for our estimations.\footnote{%\XZ{added}
More refined analysis is given in \cite{Komatsu:2024ydh}, where a similar assumption on matrix model side is also discussed.}

Now we analyze the curvature and the dilaton for different values of the dimensionless parameter $\rho_s/z_s$. In fact, to connect with our matrix model discussion in \cite{Komatsu:2024ydh}, it will be more convenient to use the parameter 
\begin{equation}
    \xi \equiv \frac{3^5}{2^{20}\pi^2} \frac{Q_s}{z_s^5 \mu^5} = \frac{n_s}{\Omega^4 N_s^5},
\end{equation}
which is related to $\rho_s/z_s$.
As we show in \cite{Komatsu:2024ydh}, we can write
$
    \frac{\rho_s}{z_s} = w(\xi),
$
where $w(\xi)\sim \xi^{1/5}$ when $\rho_s/z_s \gg 1$ and $w(\xi)\sim \xi^{1/4}$ when $\rho_s/z_s \ll 1$. 
% \XZ{I think it's useful to remind the readers here that large $\r_s$ means large $Q_s$ (even better if can be more precise) because we switch between these two in the discussions.} \AV{It can be more subtle because $Q_s$ is just $n_s$ but $\rho_s$ contains the coupling $\Omega$. So we can take $n_s$ large but $\xi$ small (and then $\rho_s/z_s$ small) if $\Omega$ is very large (let's say $N_s$ fixed)}
%\JP{Changed previous sentence.}
We can now study the dilaton and curvature for different regimes of $\xi$, in regions close to the axis, the disk, and the conducting plane. In all these regions, we get 
% \XZ{In what limit do the following $\sim$ relations hold?} \AV{close to the axis, disks, and conducting plane}
% If we assume $z_s \sim \rho_s$ and $Q_s \sim z_s^5 \mu^5$ \AV{that corresponds to $\xi$ fixed, can we justify this here ?} we get 
% \JP{I suppose the final estimates are multiplied by functions $q_1(\xi)$ and $q_2(\xi)$ assuming $\xi$ is fixed. It would be instructive to determine the behavior of these functions when $\xi \to 0$ and $\xi \to \infty$.} \AV{check curvatures}
 \begin{equation}
    \begin{split}
          R \alpha' \sim \frac{\alpha' \mu}{\rho_s} \tilde{q}_R(\xi)\sim \frac{q_R(\xi)}{\lambda^{1/6}}, \qquad g_s e^\phi \sim \frac{g_s \tilde{q}_\phi(\xi)}{\mu^2 \rho_s^2}\sim \frac{\lambda^{2/3}}{N} q_\phi(\xi),
        \end{split}
        \label{R and phi}
    \end{equation}
where $\tilde{q}_{R} =  \xi^{1/6} q_R/w(\xi)$ (and similarly for $\tilde{q}_\phi$), and the functions $q_R$ and $q_\phi$ have different asymptotics in different regions, see Table \ref{tab:summary of validity analysis}. We get the stronger conditions from the tip of the disks where 
\begin{equation}
    (\rho,z) \to (\rho_s,z_s) : \qquad q_R(\xi) \approx \begin{cases*}
    \xi^{-2/15}& if \  $\xi \gg 1 $ \\
       \xi^{-1/12}    & if \   $\xi \ll 1 $
    \end{cases*}\,,
    \qquad 
    q_\phi(\xi) \approx \begin{cases*}
    \xi^{-4/15} & if \  $\xi \gg 1 $ \\
       \xi^{1/12}    & if \   $\xi \ll 1 $
    \end{cases*}\,.
    \label{eq:qxi}
\end{equation}
In fact when $\xi \gg 1$ we get the same conditions in the other regions as well. %\AM{From \eqref{curvature dilaton infinity} with $\xi \gg 1$, using $z_s \sim \mu \alpha' N_s$ and $N_s \sim \lambda^{1/6} \xi^{-1/6}$ I got $\rho_s^2 \sim \xi^{2/5} z_s^2 \sim \xi^{2/5} \mu^2 \alpha'^2 \lambda^{1/3} \xi^{-1/3} $ which means $\alpha' R \sim \lambda^{-1/6} \xi^{1/15}$ instead of $\xi^{-2/15}$. Where am I wrong?} \AV{I think that is true, it adds the condition $x^{(s)} N_s \gg 1$ when $\xi \gg 1$. I updated below.}  
However when $\xi \ll 1$ we get larger powers in the other regions. For example on the axis we have $q_\phi \sim \xi^{4/3}$ which is much less than $\xi^{1/12}$ when $\xi$ is small. We also find that those conditions are stronger than imposing small string coupling in \eqref{curvature dilaton infinity} at $r\sim \operatorname{max}(\rho_s,z_s)$. However the asymptotic curvature gives a stronger condition, where we find $q_R \sim \xi^{1/15}$ when $\xi \gg 1$ and $q_R \sim\xi^{-1/3}$ when $\xi \ll 1$. For the former, it turns our that the condition is automatic when $\xi \gg 1$. However for the latter we get an extra constraint.

In addition we also have the constraint $N_{D5} \gg 1$, which is required for the electrostatic problem to not be overconstrained as we discussed just below \eqref{eq:ND5}\footnote{\textbf{Note added (April 2025)} : We thank Sean Hartnoll and Jun Liu for pointing this out \cite{Hartnoll:2025ecj}.}. We find 
\begin{equation}
    N_{D5} \sim \frac{N}{\lambda^{2/3}}q_{D5}(\xi),
\end{equation}
where $q_{D5}(\xi) \sim \xi^{1/15}$ when $\xi \gg 1$ and $q_{D5}(\xi) \sim \xi^{-1/3}$ when $\xi \ll 1$. The constraint when $\xi \gg 1$ is not implied by the previous ones.

In terms of the parameters $(x^{(s)},n_s,N_s)$, where $x^{(s)} = \frac{1}{2\pi \mu \alpha'} \rho_s$ \cite{Komatsu:2024ydh}, the supergravity approximation is valid for
    \begin{equation}
    n_s \gg (x^{(s)})^3 \gg N_s x^{(s)}.
    \label{eq:valSUGRA}
\end{equation}
In particular, the irreducible vacuum with $n_s=1$ is not well described by supergravity.
It is interesting to compare these conditions with the ones obtained in \cite{Komatsu:2024ydh} for the derivation of the electrostatic problem  from the exact localization computation.
There, we needed
\begin{align}
    n_s \gg (x^{(s)})^2 \log x^{(s)} \gg 1\,,
\end{align}
which is a weaker condition.

%\JP{Added this speculative sentence.}

% If instead we expand near the center of the disks, origin, $z$-axis or conducting plane, the curvature and dilaton take a form similar to \eqref{R and phi}, but with different asymptotics for the functions $q_R$ and $q_\phi$ 
% \AV{below the dilatons are fine but I need to double check curvature which takes a lot of time on the computer, but from the eq above it seems that $\xi \to 0$ will be a bad limit since the curvature blows up (that's the irrep)}
% \JP{What is $\delta \rho$ and $\delta z$? If it is the distance then it means the curvature and the dilaton diverges?} \AV{not sure about the expansion around $(0,0)$, using $\rho \to 0$ with $z$ fixed and then $z \to 0$ seems different than $z \to 0$ with $\rho$ fixed and then $\rho \to 0$...}

\begin{table}[]
    % \centering
    % \begin{tabular}{c|c|c|c}
    %     Regions in Fig.\ref{fig:sugra_regime} & $q_R(\xi)$ & $q_\f(\xi)$  \\
    %      & 
    % \end{tabular}

\begin{center}
\begin{tabular}{|c||c|c|c|c|c|}
    \hline
    \multirow{2}{*}{Regions in Fig.\ref{fig:sugra_regime}}
    & \multicolumn{2}{|c|}{$q_{R}(\xi)$}  
    & \multicolumn{2}{|c|}{$q_{\f}(\xi)$}\\
    \cline{2-5}
     & $\xi\gg1$ & $\xi\ll1$ & $\xi\gg1$ & $\xi\ll1$
    \\ \hline\hline
    \circled{1} & $\xi^{1/15}$ & $\xi^{-1/3}$ & $\xi^{-4/15}$ & $\xi^{4/3}$ 
    \bigstrut \\ \hline
    \circled{2},\circled{4} & $\xi^{-2/15}$ & $\xi^{-1/12}$ & $\xi^{-4/15}$ & $\xi^{1/12}$
    \bigstrut \\ \hline
    \circled{3} & $\xi^{-2/15}$ & $\xi^{-1/12}$ & $\xi^{-4/15}$ & $\xi^{1/6}$
    \bigstrut \\ \hline
\circled{5},\circled{6},\circled{7} & $\xi^{-2/15}$ & $\xi^{2/3}$ & $\xi^{-4/15}$ & $\xi^{4/3}$ 
    \bigstrut \\ \hline
\end{tabular}
\end{center}
\caption{Asymptotic behavior of the scalar curvature and the dilation in different regions in figure \ref{fig:sugra_regime}. For region \circled{1} we have used \eqref{curvature dilaton infinity} with $r^2 = \r_s^2 + z_s^2$. The relevant entries that lead to the final result \eqref{eq:valSUGRA} are from regions \circled{1} and \circled{2}.}
\label{tab:summary of validity analysis}
\end{table}

\subsection{Free energy}\label{subsec:freeenergy}
With the geometries at hand, we want to compute the on-shell action associated to each solution. This is expected to match the free energy of the matrix model in the large $N$ limit. %\XZ{added:} In this paper we only consider the case where on the matrix model side the saddle points $J^k$ consist of irreps with large dimension and small degeneracies, and on the gravity side the distances between conducting disks $d_i$ are all large. \AV{Our formula \eqref{eq:free_energy} does not depend on this approximation, therefore I'm not sure if we should say that we "only" consider the case of large irreps.} We leave a more complete analysis to the future work.

As shown in \cite{Okuda:2008px} (see also \cite{Gutperle_2017} for a similar computation), in type IIB supergravity with vanishing 5-form flux the on-shell action is the boundary term
\begin{equation}
\label{eq: on shell action IIB}
    S_E = \frac{1}{2\kappa^2}\int d\left(- \frac{1}{4} M_{ij} \mathcal{C}_2^i \wedge *\mathcal{F}_3^j \right),
\end{equation}
where
\begin{equation}
    M_{ij} = e^{\phi} \begin{pmatrix}
        \chi^2 + e^{-2\phi} & -\chi \\ -\chi & 1 
    \end{pmatrix}, \qquad \mathcal{C}_2 = \begin{pmatrix}
        B_2 \\ C_2
    \end{pmatrix}, \qquad \mathcal{F}_3 = d \mathcal{C}_2.
\end{equation}
This integral gets contributions from the different cycles in the geometry. This can be understood as follows. Pick the contour going around the disks and along the $z$ axis, as shown on Figure \ref{fig:integration_contour}. The interior is a patch covering the entire spacetime minus some part of measure zero. However all the cycles in that patch are contractible. We can therefore integrate over that patch and use Stokes theorem to evaluate the on-shell action. %\footnote{The equivalent statement in electromagnetism would be to integrate the 2-form flux over the full sphere with the north pole removed. Using Stokes theorem we integrate the potential around a small circle around the north pole, which picks up the contribution from the magnetic monopole.}. 
We find that the integrand vanishes on the $z$ axis. 
Therefore we are left with the contributions from going around the disks, which give
\begin{equation}
\begin{split}
    2\kappa^2 S_E &= -\frac{5120}{3\mu^{10}}\Omega_6 \Omega_2 \sum_{\text{disks}\ s} \int d\rho \left(  \rho^3 V_s \dot{V}' - \frac{1}{2} z_s \rho^3 \partial_\rho( V'^2) \right) \\
\end{split}
\end{equation}
where the domain of integration is above and below the disk. The second term vanishes since $V'^2$ is continuous across the disk. Evaluating and using $2 \kappa^2 \equiv 16\pi G_N^{(10)} = (2\pi)^7 \alpha'^4 g_s^2$ we get the answer 
\begin{equation}
\label{eq:free_energy}
    S_E = \frac{2^7}{3 \pi^4 g_s^2 \alpha'^4\mu^{10}} \sum_{s=1}^q Q_s V_s.
\end{equation}
We then find that the free energy of the supergravity solution corresponds to the electrostatic energy of the disk configuration\footnote{This is not entirely exact because of the presence of the background electric field.}.

\begin{figure}[h]
\centering
    \begin{tikzpicture}[domain=0:4]

  \draw[->] (0,0) -- (4.2,0) node[right] {$\rho$};
  \draw[->] (0,0) -- (0,4.2) node[above] {$z$};
  \draw[color=blue]   (0,0) -- (4,0);
  \draw[color=blue]   (0,1) -- (2,1) ;
  \draw[color=blue]   (0,2) -- (1,2);
  \draw[color=blue]   (0,3) -- (1.5,3);

  \draw[color=darkgreen,->]   (0.2,0.2) -- (4,0.2) arc(0:45:3.8);
  \draw[color=darkgreen]   (0.2,4) arc(90:45:3.8);
  \draw[color=darkgreen] (.2,4)--(.2,3.2);
  \draw[color=darkgreen] (.2,2.8) -- (.2,2.2) -- (1,2.2) arc(90:-90:.2) -- (.2,1.8);
  
  \draw[color=darkgreen]   (0.2,3.2) -- (1.5,3.2);
  \draw[color=darkgreen] (1.5,2.8) arc(-90:90:0.2);
  \draw[color=darkgreen]   (0.2,2.8) -- (1.5,2.8);
  
\draw[color=darkgreen] (0.2,1.8) -- (0.2,1.2) -- (2,1.2) arc(90:-90:0.2) --(0.2,0.8)--(0.2,0.2);
\end{tikzpicture}
    \caption{Integration contour for the on-shell action. The interior is a patch covering the entire spacetime minus regions of measure zero. All cycles in that patch are contractible.}
    \label{fig:integration_contour}
\end{figure}

% \begin{figure}[h]
% \centering
%     \begin{tikzpicture}[domain=0:4]

%   \draw[->] (0,0) -- (4.2,0) node[right] {$\rho$};
%   \draw[->] (0,0) -- (0,4.2) node[above] {$z$};
%   \draw[color=blue]   (0,0) -- (4,0);
%   \draw[color=blue]   (0,1) -- (2,1) ;
%   \draw[color=blue]   (0,2) -- (1,2);
%   \draw[color=blue]   (0,3) -- (1.5,3);

%   \draw[thin,color=green,->]   (0.2,0.2) -- (4,0.2) arc(0:45:3.8);
%   \draw[thin,color=green]   (0.2,4) arc(90:45:3.8);
%   \draw[thin,color=green] (.2,4)--(.2,3.2);
%   \draw[thin,color=green] (.2,2.8) -- (.2,2.2) -- (1,2.2) arc(90:-90:.2) -- (.2,1.8);
  
%   \draw[thin,color=green]   (0.2,3.2) -- (1.5,3.2);
%   \draw[thin,color=green] (1.5,2.8) arc(-90:90:0.2);
%   \draw[thin,color=green]   (0.2,2.8) -- (1.5,2.8);
  
% \draw[thin,color=green] (0.2,1.8) -- (0.2,1.2) -- (2,1.2) arc(90:-90:0.2) --(0.2,0.8)--(0.2,0.2);
% \end{tikzpicture}
%     \caption{Integration contour for the on-shell action. The interior is a patch covering the entire spacetime minus regions of measure zero. All cycles in that patch are contractible.}
%     \label{fig:integration_contour}
% \end{figure}

The explicit expression of the gravitational on-shell action requires computing the potentials $V_s$ by solving an electrostatic problem. Let us focus on the particular limit when $d_s \gg 1$. In this case we can consider only one disk at a time in the background potential. Its charge density can be computed explicitly as we show in appendix \ref{app:electrostatics}. Using the constraint that it vanishes at the tip we find 
\begin{equation}
    V_s = -\mu^5\frac{z_s^3}{2^7} +\mathcal{O}(\sqrt{Q_s z_s}),
\end{equation}
which is just the contribution from the background potential evaluated at the center of the disk.
Inserting all the factors, together with $g_{YM}^2 = \frac{g_s}{(2\pi)^3 \alpha'^2}$ the gravitational on-shell action gives
\begin{equation}
    S_E = - \frac{\mu^4}{g_{YM}^2} \frac{9}{2^{17}} \sum_s n_s N_s^3 + \mathcal{O}( \sqrt{n_s N_s})\,.
\label{eq:free energy in large d_i limit}
\end{equation}

It is enlightening to compare this result to expectations from the matrix model. Because of the identification \eqref{eq:parameter identification}, the limit $d_s\gg1$ corresponds to the scenario in the matrix model side where the saddle points \eqref{eq:fuzzy_spheres_saddles_in_J^k} consist of $SU(2)$ irreducible representations with large matrix dimension and small degeneracy. This is the limit where the fuzzy spheres become almost classical and the quantum fluctuations are expected to be negligible.\footnote{%\XZ{added:}  
In the companion paper \cite{Komatsu:2024ydh} we show that, using supersymmetry localization, the \emph{full} free energy of the matrix integral has exactly the same expression as \eqref{IKKT_saddle_action} in the same limit, despite the computation being highly non-trivial.}  Taking only the contributions from the saddles, we find that the value of the action is
\begin{equation}
\label{IKKT_saddle_action}
    S_{\Omega}^{\rm saddle} = -\Omega^4 \frac{9}{2^{15}} \sum_{s=1}^q n_s N_s (N_s^2-1),
\end{equation}
where $n_s$ is the quantized charge $Q_s$ and $N_s$ is the quantized height $z_s$. We see that the functional forms of \eqref{eq:free energy in large d_i limit} and \eqref{IKKT_saddle_action} are identical (at large $N_s$).
%\JP{I thought it was $d_i$}. \AV{$N_{NS5,i}$ is $d_i$, but $z_i$ is obtained by summing $N_i \sim z_i =\sum_{j<i} d_i \sim \sum_{j<i}N_{NS5,i} $}. 
%Using the relation \eqref{eq:Omega_mu} which will be shortly discussed, this result matches with \eqref{eq:free energy in large d_i limit} up to a factor of $4$.

There is still a leftover piece in the evaluation of the on-shell action, namely the contribution from the asymptotic boundary of spacetime. This gives both a divergent term and a finite term. %\XZ{since the spacetime is asymptotically flat (correct?)} \AV{I think it could be divergent also in a different asymptotic geometry}\XZ{But in our case (geometry dual to polarized IKKT) it's asymptotically flat right?}. 
The finite term comes from the octopole of the electrostatic configuration $q_3$. To see this, we expand 
\begin{equation}
    V(\rho,z) = -\eta \mu^3 z+\frac{\mu^5}{2^7}(z \rho^2-z^3) + \frac{P}{2\pi^2} \frac{z}{(z^2+\rho^2)^2}+ \frac{q_3}{\pi^2} \frac{z(z^2-\rho^2)}{(z^2+\rho^2)^4}+...
\end{equation}
and evaluate the contribution from the arc at $\rho^2+z^2 = R^2$:
\begin{equation}
\label{eq:S inf}
    2 \kappa^2 S_E^{(\infty)} =  \Omega_2 \Omega_6 \frac{2^4}{\mu^5} \left(\frac{P  R^2}{\pi} + \frac{5 q_3}{3 \pi} \right) + \mathcal{O}(R^{-2}).
\end{equation}
Since we are considering this spacetime boundary we should also add the Gibbons-Hawking term, for which we get 
\begin{equation}
\label{eq: S GH}
    \kappa^2 S^{(GH)} =  -\int_{\partial M} \sqrt{h}K = \Omega_2 \Omega_6  \left(\frac{45 \pi R^8 }{256}- \frac{36 P R^2}{ \pi \mu^5} 
    -4 \frac{q_3}{  \pi \mu^5} \right)+ \mathcal{O}(R^{-2}).
\end{equation}
As can be seen from these expressions, the Gibbons-Hawking term alone cannot remove the diverging terms. This indicates that one needs to supplement it with suitable counter terms. This procedure will in principle change the finite term. We leave  for the future the determination of the counter terms and the precise match with the matrix model side.

%\SK{OK, let's keep it}
For now, note that in the limit $d_s \gg 1$, we can compute the octopole $q_3$ as shown in appendix \ref{app:electrostatics}, where we find 
\begin{equation}
    q_3 = 2\sum_s Q_s d_s^3 + \mathcal{O}\left(\sqrt{N_s n_s^3}\right).
\end{equation}
The finite pieces of the contributions \eqref{eq:S inf} and \eqref{eq: S GH} are then 
\begin{equation}
    S^{(\infty)}_{\text{finite}} = \frac{\mu^4}{g_{YM}^2}\frac{3}{2^{13}} \sum_s n_s N_s^3+ \mathcal{O}\left(\sqrt{N_s n_s^3}\right), \qquad  S^{(GH)}_{\text{finite}} = -\frac{\mu^4}{g_{YM}^2}\frac{9}{5 \cdot 2^{14}} \sum_s n_s N_s^3+ \mathcal{O}\left(\sqrt{N_s n_s^3}\right),
\end{equation}
and we see that they again take the expected functional form in terms of $n_i$ and $N_i$. Adding those contributions together with the bulk term \eqref{eq:free energy in large d_i limit} does not reproduce the numerical prefactor of \eqref{IKKT_saddle_action} under the identification of parameters \eqref{eq:Omega_mu} that we discuss in the next section, suggesting that the counterterms modify the finite contributions.

 \paragraph{Scaling of the free energy.} 
 %\AV{added} \JP{edited}
% There are two scaling transformations that are symmetries of the solutions and under which the full potential (background $+$ disks) transform homogeneously 
% \begin{equation}
% \begin{split}
%     (\rho,z) &\to (\lambda_1 \rho, \lambda_1 z), \qquad \mu \to \lambda_1^{-1}\,\mu,\qquad V \to \lambda_1^{-2}\, V, \qquad S \to \lambda_1^8 S, \\
%     Q_s &\to \lambda_2 Q_s, \qquad \mu \to \lambda_2^{1/5} \mu, \qquad V \to \lambda_2 V, \qquad S \to S
% \end{split}
% \end{equation}
% Note that both leave invariant the quantity
% \begin{equation}
%     \xi \equiv \frac{3^5}{2^{20}\pi^2} \frac{Q_s}{z_s^5 \mu^5} = \frac{n_s}{\Omega^4 N_s^5}.
% \end{equation}
The on-shell action depends on 3 parameters: the positions of the disks $z_s$, their charges $Q_s$ and the parameter $\mu$. To consider dimensionless quantities we write $S= S(Q_s, \mu z_s, \mu \sqrt{\alpha'})$.
We will use two symmetries. Firstly, notice that
\begin{equation}
   S(Q_s, \mu z_s, \mu \sqrt{\alpha'}) = \lambda_1^{-8} S( Q_s, \mu z_s, \lambda_1^{-1} \mu \sqrt{\alpha'}) \,.
\end{equation}
This follows from the transformation
\begin{align}
    (\rho,z) &\to (\lambda_1 \rho, \lambda_1 z), \qquad \mu \to \lambda_1^{-1}\,\mu,
\end{align}
which implies $V \to \lambda_1^{-2}\, V$ and  $Q_s$ remains fixed.
Secondly, notice that if we  rescale 
$\mu \to \lambda_2^{1/5} \mu$ and $Q_s \to \lambda_2 Q_s$ keeping $z_s$ fixed, then $V \to \lambda_2 V$ 
% and  \AV{I think the rescaling of $Q$ does not follow from the one of $\mu$ and has to be considered independently}
. This leads to 
\begin{equation}
   S(\lambda_2 Q_s, \lambda_2^{1/5}\mu z_s, \lambda_2^{1/5}\mu \sqrt{\alpha'}) =  S( Q_s, \mu z_s,  \mu \sqrt{\alpha'}) \,.
\end{equation}
Note that both transformations leave invariant the quantity\footnote{In principle we should have a different $\xi$ for each disk, i.e. $\xi = \xi_s$, but the scaling argument follows through.}
\begin{equation}
    \xi \equiv \frac{3^5}{2^{20}\pi^2} \frac{Q_s}{z_s^5 \mu^5} = \frac{n_s}{\Omega^4 N_s^5}.
\end{equation}
Using the symmetries above we have the scaling relation
\begin{equation}
   S(Q_s, \mu z_s, \mu \sqrt{\alpha'}) = \lambda_1^{-8} S( Q_s, \mu z_s, \lambda_1^{-1} \mu \sqrt{\alpha'}) = \lambda_1^{-8} S \left(\lambda_2 Q_s, \lambda_2^{1/5} \mu z_s, \lambda_2^{1/5}\lambda_1^{-1} \mu \sqrt{\alpha'} \right).
\end{equation}
Using $\lambda_1= \sqrt{\alpha'} \mu Q_s^{-1/5}$ and $\lambda_2= Q_s^{-1}$ we get 
\begin{equation}
    S(Q_s, \mu z_s, \mu \sqrt{\alpha'})  = \frac{Q_s^{8/5}}{\alpha'^4 \mu^8} S\left(1, z_s \mu Q_s^{-1/5}, 1 \right) = \frac{Q_s^{8/5}}{\alpha'^4 \mu^8} S\left(1, \xi^{-1/5}, 1 \right).
\end{equation}
Using the relations \eqref{eq:electrostatic parameter matching} we get 
\begin{equation}
    \frac{Q_s^{8/5}}{\alpha'^4 \mu^8} \propto \xi^{4/15} N^{4/3} \Omega^{8/3},
\end{equation}
such that 
\begin{equation}
    S(Q_s, \mu z_s, \mu \sqrt{\alpha'}) = \frac{N^2}{\lambda^{2/3}}H(\xi),
\end{equation}
where $\lambda \equiv N/\Omega^4 = N g_{YM}^2/\mu^4$ is the dimensionless 't Hooft coupling. This is the predicted scaling that has been discussed from scaling similarity \cite{Bobev:2019bvq,Biggs:2023sqw,Bobev:2024gqg}.
\subsection{Polarized probe $D1$ brane}
\label{sec: probe D1}
%\AV{Maybe this should not be in section 3 which is "backreacted geometry".Should we change the name of the section or create a new one ?}
Here we study the system from a probe brane perspective. Starting from a stack of $N$ $Dp$ branes, its non-abelian DBI action contains couplings to higher form gauge fields with respect to which a single $Dp$ brane is neutral. In the particular case of $p=0$, a 4-form RR flux polarizes the $D0$ branes into a $D2$ brane, which is known as the Myers effect \cite{Myers:1999ps}. See \cite{Lin:2004kw} for an analysis of this effect in the BMN model. Here we study a similar system where $N$ $D$-instantons polarize into a $D1$ brane under the effect of an external constant RR flux. A similar analysis with an external NSNS flux was conducted in \cite{Hartnoll:2024csr}.

The DBI action for a single $D1$ brane reads
\begin{equation}
\begin{split}
    S_{D1} &= -\frac{1}{2 \pi \alpha' g_s} \int d^2 \sigma e^{-\phi} \sqrt{- \operatorname{det}\left(g^{(s)}_{\alpha \beta}+B_{\alpha \beta}+2\pi \alpha' F_{\alpha \beta}\right)}\\ &+\frac{1}{2 \pi \alpha' g_s} \int \left(\chi\left(B_2 + 2\pi \alpha'F_2\right) -C_2 \right),
\end{split}
\end{equation}
where $g,B_2,\chi$ and $C_2$ are respectively the pullbacks of the string frame metric $g^{(s)} \equiv e^{\phi/2} g$, NSNS potential, axion and RR potential, while $F_2$ is the worldvolume $U(1)$ field strength. For a bound state of $N$ $D$-instantons it is given by $F_2 = \frac{N}{2} d \Omega_2$ \cite{Myers:1999ps,Hartnoll:2024csr}.  
This is the Lorentzian action evaluated with purely imaginary RR fields and a metric with Euclidean signature. We get a real Euclidean action by sending $S \to i S$.

 Following our physical picture, we would like to start with $D$-instantons in flat space, and add a flux, which would define the background fields. However, flat space and constant flux is not a supergravity solution unless we also add a non-trivial axi-dilaton. We can study what this solution can be by using our backreacted geometries and take a \emph{probe} limit where the backreaction goes to zero. Therefore we expand
\begin{equation}
    V(\rho,z) = -\eta z \mu^3  +\frac{\mu^5}{2^7}(
z \rho^2-z^3)+ \epsilon f(\rho,z),
\end{equation}
and we take the limit $\epsilon \to 0$. 
We find
\begin{equation}
    \begin{split}
        ds^2 &= dz^2 + d\rho^2 + z^2 d\Omega_2^2 + \rho^2 d\Omega_6^2 + \mathcal{O}(\epsilon), \\
        C_2 &= -i\frac{\mu}{3} z^3\wedge d\Omega_2 + \mathcal{O}(\epsilon), \qquad B_2 =\mathcal{O}(\epsilon)  \\ i \bar{\tau} &\equiv 
        e^{-\phi}+i \chi = \frac{\mu^2}{64}(3z^2+\rho^2) -\eta + \mathcal{O}(\epsilon),
    \end{split}
    \label{eq:probe_background}
\end{equation}
where we only displayed a particular combination of the axion and dilaton that we will shortly justify. Individually we find that the $e^\phi$ 
%\XZ{should be $\f$?} \AV{ I always call $"e^\phi"$ the dilaton, which is what I think is common in string literature...} 
goes to zero and the axion diverges, leaving this configuration finite, 
%\AV{added comment about $\tau$} 
whereas the other combination $i \tau \equiv e^{-\phi}-i \chi \sim \frac{1}{\epsilon}$ diverges. %Note that it goes to zero when $\mu \to 0$ as expected.
We also note again that $3z^2+\rho^2$ is exactly the bosonic part of the mass-deformed IKKT matrix model.

Let us now go back to the DBI action. 
%\AV{rephrased} 
We will evaluate it in the constant RR flux background \eqref{eq:probe_background} first without specifying the axi-dilaton.
The embedding of the $D1$ is chosen to be localized in the $S^2$, at 
%\AV{added $\rho=0$} 
$\rho=0$ and spherically symmetric with radius $r$. Since everything is spherically symmetric we can integrate over the angles and we get 
\begin{equation}
    S_{D1} =  \frac{2}{g_s \alpha'} e^{-\phi} \sqrt{\pi^2 \alpha'^2 N^2+ e^{\phi}r^4}+ \frac{2\pi N }{g_s} (i\chi) %+ \frac{2i C_2}{g_s \alpha'}
    - \frac{2 r^3 \mu}{3 \alpha' g_s}.
\end{equation}
At large $N$ we get 
\begin{equation}
   S_{D1} =  \frac{2 \pi N}{g_s}\left(e^{-\phi}+i \chi \right)  + \frac{r^4}{g_s \pi \alpha'^2 N} 
   - \frac{2 r^3 \mu}{3 \alpha' g_s}
   %+ \frac{2i C_2}{g_s \alpha'} 
   + \mathcal{O}(N^{-2}),
\end{equation}
where we see the relevant axi-dilaton appearing. We can now plug in our background \eqref{eq:probe_background} and we get
\begin{equation}
    S_{D1} = \frac{r^4}{N g_s \pi \alpha'^2}- \frac{2 r^3 \mu}{3 \alpha' g_s}+ \frac{3 N \pi r^2 \mu^2}{32 g_s}-\frac{2\pi N \eta}{g_s} +\mathcal{O}(N^{-2}).
\end{equation}
%\AV{Comment about $\xi$ and the trace part of the matrix model ? Also note that from \eqref{eq:probe_background} we get $i \bar{\tau}(r=0) = i\chi+ e^{-\phi} |_{r=0} = -\xi  $ and $S_{D1}(r=0)=-\frac{2\pi N \xi}{g_s}=\frac{ 2\pi i \bar{\tau}}{g_s}(r=0)$, which is consistent with the analysis of \cite{Green:1997tv}}\SK{Added. See below (3.60).}
The allowed configurations of the $D1$ are found by looking at extrema of the action $S_{D1}'(r)=0$. We get the solutions 
\begin{equation}
    r=0, \qquad r= \frac{1}{8}N \pi \mu \alpha', \qquad  r= \frac{3}{8}N \pi \mu \alpha',
\end{equation}
which are the same roots as the ones for the fuzzy sphere saddles of the polarized IKKT integral \eqref{eq:fuzzy_spheres_saddles}. The global minimum occurs for the third solution, and the on-shell action reads
\begin{equation}
    S_{D1} = -\frac{2\pi N\eta}{g_s}-\frac{9}{2^{15}}N^3 \frac{\mu^4}{g_{YM}^2}= \frac{2\pi iN\bar{\tau}(r=0)}{g_s}-\frac{9}{2^{15}}N^3 \frac{\mu^4}{g_{YM}^2}.
    \label{eq:onshellD1}
\end{equation}
In the second equality, we used the relation $i \bar{\tau}(r=0) = i\chi+ e^{-\phi} |_{r=0} = -\eta  $, obtained from \eqref{eq:probe_background}. 

%\SK{Added:}
The first term in \eqref{eq:onshellD1} can be identified with the on-shell action of $N$ $D$-instantons at $r=0$ (cf. \cite{Green:1997tv}). As discussed in \cite[Sec.4.3]{Hartnoll:2024csr},  this can be reproduced on the matrix model side by adding a constant term $ \frac{2\pi \bar{\tau}(r=0)}{g_s}{\rm Tr}(\mathbb{1})$ to the action. On the other hand, the second term is a contribution from polarizing $D$-instantons into $D1$ and is dominant when $N\gg 1$. This second term should be compared with the (irrep.) fuzzy sphere on-shell action on the matrix model
\begin{align}
    % S_{D1}= - \frac{9}{2^{15}} N^3 \frac{\mu^4}{g_{YM}^2}, \qquad 
    S_{\W}^{{\rm fuzzy\ sphere}} = - \frac{9}{2^{15}} N^3 \Omega^4.
\end{align}
The comparison leads to a matching of the parameters on both sides,
\begin{equation}
    \Omega = \frac{\mu}{\sqrt{g_{YM}}}.
    \label{eq:Omega_mu}
\end{equation}

To further test the correspondence between the matrix model side and the gravity side, we consider the size of the (irrep.) fuzzy spheres on the matrix side using 
\begin{equation}
    r_{\text{fuzzy sphere}}^2=(2\pi \alpha')^2 \frac{g_{YM}}{N}\operatorname{Tr} X^2 =(2\pi \alpha')^2  g_{YM}\frac{3^2}{8^2} \Omega^2 \frac{N^2-1}{4} \simeq \left(\frac{3}{8} \pi \alpha' \sqrt{g_{YM}} \Omega N\right)^2.
\end{equation}
This exactly matches the radius of the probe $D1$ brane 
\begin{equation}
    r_{D1} = \frac{3}{8} N \pi \mu \alpha',
\end{equation}
if we again use the matching conditions \eqref{eq:Omega_mu}. We can also look for an analog in the backreacted geometries. There we can compute the height of the disk using the asymptotic flat metric. That gives%\footnote{Here we are using the asymptotic metric before the rescaling mentionned just above \eqref{eq:probe_background}.}
\begin{equation}
    r_{\text{disk}}=\int_0^{d} \sqrt{g_{zz}}dz = d   = \frac{3}{8} N\pi \mu \alpha',
\end{equation}
again matching the previous results.
We thus find a perfect match among the three computations, 
\begin{equation}
    r_{\text{fuzzy sphere}}= r_{D1}=r_{\text{disk}}\,.
\end{equation}

\paragraph{Comparison with the probe brane analysis in \cite{Hartnoll:2024csr}.} Our identification of parameters \eqref{eq:Omega_mu} differs from the one obtained in \cite{Hartnoll:2024csr} by a factor of 2. However the physical systems we analysed are different, hence it is not a direct contradiction. As we explained in \ref{sec:asymptotics}, our background \eqref{eq:probe_background} is related to their background \eqref{eq:cavity} by S-duality. However we are both studying a $D1$ brane in our respective backgrounds, whose S-dual is an $F1$ string. Hence the two systems are not equivalent and correspond to different physics. 

Let us argue for our choice. We will give two reasons. First, studying the polarization of $Dp$ branes due to an external RR flux (rather than NSNS) seemed more natural in view of previous works \cite{Myers:1999ps,Lin:2004kw}. Second,
%\AV{added more importantly} 
and more importantly, in our backreacted geometries we could measure the number of $F1$ strings \eqref{eq:F1zero} and found zero. Therefore, even if it is perfectly fine to study the dynamics of a probe $F1$ on this background, we cannot argue that the backreacted geometry corresponds to the backreaction of this $F1$ string. If we S-dualize \eqref{eq:sugra_solution} so that the asymptotic background \eqref{eq:cavity} is identical to the one considered in \cite{Hartnoll:2024csr}, we find that the number of $D1$ branes is zero. Thus it is more natural to consider a probe $F1$ string as opposed to a probe $D1$ brane as was done in \cite{Hartnoll:2024csr}.

\section{Discussion}
\label{sec:discussion}

\paragraph{Summary and main results.}

In this paper we studied Euclidean type IIB geometries dual to the mass-deformed IKKT matrix model, which are closely related to Lin-Maldacena geometries. Indeed they are smooth and without horizons, and can be formulated in terms of a four-dimensional electrostatic potential $V(\rho,z)$. Asymptotically the metric describes flat Euclidean space (in the Einstein frame) with an axi-dilaton corresponding to the $D$-instanton solution \cite{Gibbons:1995vg,Ooguri:1998pf}. In addition there is a constant 3-form Ramond-Ramond flux. This is consistent with the picture that those geometries correspond to the backreaction of $N$ $D$-instantons polarizing into a $D1$ brane \cite{Hartnoll:2024csr}. Those geometries are in one-to-one correspondence with the fuzzy sphere vacua of the polarized IKKT matrix model, as we showed explicitly by quantizing the different fluxes and relating their quantum numbers to the dimensions and degeneracies of the $SU(2)$ irreducible representations. %Finally we derived a formula for the free energy of the solutions, and matched it with the value of the IKKT integrant at the saddles in the limit of large $SU(2)$ representation.

%\paragraph{Free energy} \AV{remove this ?} Our result for the on-shell action of the supergravity solution \eqref{eq:free_energy} does not quite match the expectation from the matrix model \eqref{IKKT_saddle_action}. The form of the equation is correct but it is missing a factor of 4. However the on-shell action \eqref{eq: on shell action IIB} also receives divergent and finite contributions from infinity. In addition there is the Gibbons-Hawking term, that also has both divergent and finite contributions. It turns out that all finite contributions are proportional to the multipole $q_3$ of the electrostatic configuration. Nevertheless, adding those contributions does not give the correct factor to match the matrix model on-shell action. We still have to carefully tame the divergences by introducing a counterterm action, which will in principle change the finite term.

In the companion paper \cite{Komatsu:2024ydh} we compute the partition function and some correlation functions of the polarized IKKT model using supersymmetric localization. Our preliminary results suggest that, in the large $N$ limit, the localization equations are identical to the electrostatic problem that determines the supergravity solution, 
%\XZ{added:} 
similar to the case for the BMN model \cite{Asano:2014eca}. In particular we can write the partition function in terms of the electrostatic variables, where it also receives two contributions, one from the electrostatic energy $Q_s V_s$, and one from the octopole $q_3$, matching respectively the bulk term and the boundary term of the supergravity on-shell action. The detailed comparison with the supergravity answer is in progress and we hope to report it soon.

\paragraph{Future directions.}
\begin{itemize}
    \item In this paper, we constructed backreacted geometries but did not evaluate their on-shell actions. As discussed in section \ref{subsec:freeenergy}, doing so requires determining the boundary counterterms. This can be determined by requiring supersymmetry to be restored, see e.g. \cite{Freedman:2016yue}.
    \item As mentioned above, the supersymmetric localization allows us to compute certain correlation functions as well. As in the usual holography, operators in the matrix model should be dual to some asymptotic behaviour of the supergravity fields. It would be enlightening to derive the dictionary carefully and perform quantitative comparison of correlation functions on both sides.
   \item In this paper, we focused on solutions without D7-brane charges. It would be interesting to generalize our construction to include D7-branes (i.e.~nontrivial SL(2,$\mathbb{C}$) monodromies). Similar solutions dual to five-dimensional superconformal field theories were constructed in \cite{DHoker:2017zwj} and it should be possible to perform suitable analytic continuation to obtain geometries with $SO(7)\times SO(3)$ isometry. 
   \item A closely related question would be to identify the matrix-model counterparts of geometries with D7-branes. One possibility may be to look for mass deformation of the $D(-1)/D7$ matrix models discussed recently in \cite{Billo:2021xzh,Aguilar-Gutierrez:2022kvk,Reymond:2024mwe} that preserve exceptional $F_4$ superalgebra.
    \item In the limit $\Omega \to 0$, one could derive an effective action for the diagonal modes by explicitly integrating out the off-diagonal modes. This has been done for the IKKT model in \cite{Aoki:1998vn}. For the BMN model, the effective Hamiltonian resulting from similar computation reduces to decoupled supersymmetric harmonic oscillators and its spectrum reproduces that of 11D (linearized) supergravity on the pp-wave background to the leading order \cite{Komatsu:2024vnb}. However, it turns out to be significantly more difficult when one tries to compute subleading corrections to the BMN effective Hamiltonian, which is a key for understanding how backreacted geometry forms through interactions of graviton gas. With the simpler matrix model one can perhaps make more progress.
   % \item Since IKKT can be derived by dimensional reduction of the super-Yang-Mills theory, it can be thought of as the high temperature limit of the BFSS model. Is there an analog for the mass-deformation ?
   %\item Our solutions asymptote to the twelve-dimensional pp-wave background. This is very similar to BFSS and BMN models whose dual geometries asymptote to M-theory backgrounds. In the BMN model, the dual M-theory  background of Lin and Maldacena \cite{Lin:2004nb} is expressed in terms of axial symmetric solutions to a three-dimensional Laplace's equation. There, the axial symmetry is tantamount to the translation symmetry along the M-theory circle. We found that the electrostatic problem for our solution is four-dimensional, also with axial symmetry, which means that the solutions are independent of 2 angular coordinates. Are they related to the translation symmetry along the F-theory torus \cite{Vafa:1996xn}?
   \item 
   In the case of the BFSS and BMN models, studying backreacted geometries \cite{Lin:2004nb, Itzhaki_1998} was important for understanding the precise relationship to M-theory and sharpening the original BFSS conjecture \cite{Banks:1996vh,Susskind:1997cw}. Therefore it would be interesting to revisit the IKKT conjecture in the light of our analysis, understand the relation to holography discussed in this paper, and possibly make the conjecture more precise. It would also be interesting to make contact with recent discussions on the (nonrelativistic) decoupling limits of string theory e.g.~\cite{Blair:2023noj,Blair:2024aqz,Lambert:2024uue,Lambert:2024yjk}. More generally it would be interesting if we can use matrix integrals to get a microscopic description of F-theory \cite{Vafa:1996xn} and extract non-perturbative observables.
\end{itemize}

\paragraph{Acknowledgements}
We thank N. Bobev, P. Bomans, F.F.Gautason,  S. Hartnoll, J. Liu, J. Matos, S. Pufu and J. Vilas Boas for discussions. This work was supported by the Simons Foundation grant 488649 (Simons Collaboration on the Nonperturbative Bootstrap) and by the Swiss National Science Foundation through the project
200020\_197160 and through the National Centre of Competence in Research SwissMAP.

\newpage
\appendix

\section{Analytic continuation of the IIB Lorentzian solution}
\label{app_analytic continuation}
In this appendix, we explain how to find the geometry \eqref{eq:sugra_solution} through analytic continuation. Our starting point is the Lorentzian solution constructed in \cite{DHoker:2016ujz}, which takes the form of a warped product $AdS_6 \times S_2 \times \Sigma_2$. The 5-form field strength vanishes, while the axion $\chi$, dilaton $\phi$, NSNS flux $H_3=dB_2$ and RR flux $F_3= dC_2-\chi B_2$ are written in terms of a complex scalar $B$ and complex 3-form $\mathcal{F}_3$ as 
\begin{equation}
    B= \frac{1+i \tau}{1-i\tau}, \qquad \tau= \chi + ie^{-\phi}, \qquad \mathcal{F}_3 = H_3 +i dC_2.
\label{eq:B tau F3}
\end{equation}
The full supergravity solution depends on only two holomorphic functions $\mathcal{A}_\pm(w)$ and reads 
\begin{equation}
\label{sugra_ansatz}
\begin{split}
    ds^2 &= f_6^2(w, \bar{w}) ds^2_{AdS_6} +f_2^2(w,\bar{w}) d\Omega_2^2 + H^2(w,\bar{w}) dw d\bar{w},\\
    \mathcal{F}_3 &= d \mathcal{C} \wedge \text{vol}_{S^{2}}, \qquad \mathcal{C} = \frac{4i}{9}\left( \frac{\bar{\partial} \bar{\mathcal{A}}_- \partial \mathcal{G}}{\kappa^2} -2R\frac{\bar{\partial} \bar{\mathcal{A}}_- \partial \mathcal{G} + \partial \mathcal{A}_+ \bar{\partial}\mathcal{G}}{(R+1)^2\kappa^2} - \bar{\mathcal{A}}_--2 \mathcal{A}_+\right), \\
    B &= \frac{\partial \mathcal{A}_+ \bar{\partial} \mathcal{G}-R \bar{\partial} \bar{\mathcal{A}}_- \partial \mathcal{G}}{R \bar{\partial} \bar{\mathcal{A}}_+ \partial \mathcal{G}-\partial \mathcal{A}_- \bar{\partial} \mathcal{G}},
\end{split}
\end{equation}
where the metric functions are 
% \XZ{function $R$ is forgotten?}\AV{thanks, added}
\begin{equation}
\label{eq:Lorentzian metric functions}
    f_6^2=  \sqrt{6 \mathcal{G} f_R}, \qquad f_2^2 = \frac{1}{9}\sqrt{\frac{6 \mathcal{G} }{f_R^3}}, \qquad H^2 = 4\kappa^2 \sqrt{\frac{f_R}{ 6\mathcal{G}}},\qquad f_R \equiv \frac{1+R}{1-R},
\end{equation}
where 
\begin{equation}
      \mathcal{G} \equiv |\mathcal{A}_+|^2-|\mathcal{A}_-|^2+\mathcal{B} +\bar{\mathcal{B}},\qquad \kappa^2 \equiv -|\partial \mathcal{A}_+|^2+|\partial \mathcal{A}_-|^2, \qquad f_R^2 =1+ \frac{2}{3} \frac{|\partial \mathcal{G}|^2}{(-\partial \bar{\partial} \mathcal{G}) \mathcal{G}},
      \label{eq:definition of G, kappa, fR}
\end{equation}
with the function $\mathcal{B}$ defined up to a constant through 
\begin{equation}
    \partial \mathcal{B} \equiv \mathcal{A}_+ \partial \mathcal{A}_--\mathcal{A}_- \partial \mathcal{A}_+.
    \label{eq:definiiton of del B}
\end{equation}
This solution can be analytic continued to a Euclidean solution by the replacement\footnote{This means that the metric 
\begin{equation}
    ds^2 = -f_6^2(w, \bar{w}) d\Omega_{6}^2 +f_2^2(w,\bar{w}) d\Omega_2^2 +H^2(w,\bar{w}) dw d\bar{w}
\end{equation}
solves the equations of motion. Indeed one can check that the corresponding changes of signs in the Ricci tensor when $AdS_6$ is replaced by $S^6$ are compensated by changes of signs of the time component when we go to Euclidean signature. This ensures that the Einstein equation is still satisfied. Then one can check that all the other equations do not depend on the sign of $f_6^2$ and are then invariant. See \cite{Corbino:2017tfl} for an explicit form of all equations of motion.
} 
\begin{equation}
    ds^2_{AdS_6} \to -d\Omega_6^2.
\end{equation}
%\JP{What do you mean?} \AV{footnote added}
We get a Euclidean solution but the metric has imaginary components. To make them real we define new coordinates $(w', \bar{w}')$ and new functions $\mathcal{G}'$, $\kappa'^2$ and $f_R'$ as 
\begin{equation}
    w = e^{-i \alpha}w', \qquad \bar{w} = e^{-i \alpha} \bar{w}', \qquad \mathcal{G} = e^{i\alpha_\mathcal{G}}\mathcal{G}',\qquad \kappa = e^{i\alpha_\kappa}\kappa',\qquad  f_R = e^{i\alpha_R}f_R',
\end{equation}
and we solve for $\alpha,\alpha_G, \alpha_\kappa$ and $\alpha_R$ so that each component of the metric is real when $G'$, $\kappa'$ and $f_R'$ are real. We find 
\begin{equation}
    \alpha = \frac{\pi}{2}, \qquad \alpha_G = \frac{3 \pi}{2},\qquad \alpha_\kappa = \frac{3 \pi}{4}, \qquad \alpha_R = \frac{\pi}{2}.
\end{equation}
This solution manages to change the sign of $f_6^2$ by going to the second sheet of the square root in \eqref{eq:Lorentzian metric functions}. We therefore have the solution 
\begin{equation}
    ds^2 = R_6^2 d \Omega_6^2 + R_2^2 d\Omega_2^2 + H^2 dw' d\bar{w}',
\end{equation}
where
\begin{equation}
\label{eq:Euclidean metric functions}
    R_6^2=  \sqrt{6 \mathcal{G}' f_R'}, \qquad R_2^2 = \frac{1}{9}\sqrt{\frac{6 \mathcal{G}' }{f_R'^3}}, \qquad H^2 = 4\kappa'^2  \sqrt{\frac{f_R'}{ 6\mathcal{G}'}}.
\end{equation}
We write the $w'$-holomorphic functions $g_\pm(w') \equiv \mathcal{A}_\pm(w(w'))$ and $h(w') \equiv \mathcal{B}(w(w'))$. Note that $\mathcal{A}_\pm(w(w'))$ and $\bar{\mathcal{A}}_\pm(\bar{w}(\bar{w}'))$ are no longer complex conjugate of each other (and similarly for $\mathcal{B}$ and $\bar{\mathcal{B}}$). Instead, we have $ \bar{\mathcal{A}}_\pm(\bar{w}(\bar{w}'))= (\mathcal{A}_\pm(-w(w')))^*$\footnote{We can see this by looking at complex conjugation of the Laurent series
% \begin{equation}
%    \bar{\mathcal{A}}(\bar{w})= \sum_n a_n^* (\bar{w}-\bar{w}_0)^n = \sum_n a_n^* (-(-i \bar{z})-\bar{w}_0)^n = \left(  \sum_n a_n (-(iz)-w_0)^n\right)^* = (\mathcal{A}(-w))^*
% \end{equation}
\begin{equation}
   \bar{\mathcal{A}}(\bar{w})= \sum_n a_n^* (\bar{w}-\bar{w}_0)^n = \sum_n a_n^* (-i \bar{w}'-\bar{w}_0)^n = \left(  \sum_n a_n (-(-iw')-w_0)^n\right)^* = (\mathcal{A}(-w))^*
\end{equation}
}. However, for $g_{\pm}(w')$ and $h(w')$ we want to keep the bar to mean complex conjugation. To this end we identify
\begin{equation}
    g_\pm(w') \equiv \mathcal{A}_\pm(w(w'))\,, \qquad \bar{g}_\pm(w')  \equiv \bar{\mathcal{A}}_\pm(-\bar{w}(\bar{w}'))\,, \qquad
    \bar{g}_\pm(\bar{w}')= (g_\pm(w'))^*\,,
\end{equation}
and similarly for $h(w')$. Now in terms of these holomorphic functions, the metric functions are 
\begin{equation}
\begin{split}
    \kappa'^2 &=  -i \left[-\partial_{w'} g_+(w') \partial_{\bar{w}'} \bar{g}_+(-\bar{w}')+ \partial_{w'} g_-(w') \bar{\partial}_{\bar{w}'} \bar{g}_-(-\bar{w}') \right] \\
    \mathcal{G}' &= i \left[g_+(w')\bar{g}_+(-\bar{w}')-g_-(w')\bar{g}_-(-\bar{w}')+ h(w') + \bar{h}(-\bar{w}') \right]\\
    \partial_{w'} h(w') &\equiv g_+(w')\partial_{w'} g_-(w')- g_-(w')  \partial_{w'} g_+(w'),
\end{split}
\label{eq:metric functions in w'}
\end{equation} 
and the defining equation for $f_R'$ is 
\begin{equation}
    f_R'^2 + 1 = \frac{2}{3} \frac{ \partial \mathcal{G}' \bar{\partial} \mathcal{G}'}{\kappa'^2 \mathcal{G}'}\,.
\end{equation}
Also note that $\partial \bar{\partial} \mathcal{G}'=\kappa'^2$, so the reality condition for the metric boils down to the construction of the real function $\mathcal{G}'$ given in \eqref{eq:metric functions in w'}. 

%\XZ{paragraph rewritten:}
The simplest solution is to have  
\al{
e^{-i\b}g_-(-w') = g_+(w') \equiv g(w')\,,
\label{eq:simple ansatz for g+}
}
with $\b$ a real constant, such that
\al{
(g_+(w')\bar{g}_+(-\bar{w}'))^* =  
\bar{g}_+(\bar{w}')  g_+(-w')
\overset{!}{=} g_-(w')\bar{g}_-(-\bar{w}')\,.
}
From \eqref{eq:simple ansatz for g+} we get automatically that $h(w') + \bar{h}(-\bar{w}')$ is imaginary (up to an inessential constant): With this identification $\del_{w'}h$ is an even function in $w'$
\al{\spl{
(\del_{w'}h)(w') =
&e^{i\b}\left[g_+(w')\del_{w'}(g_+(-w')) - g_+(-w')\del_{w'}(g_+(w'))\right]
\\
=&e^{i\b}\left[-g_+(w')(\del_{w'}g_+)(-w') - g_+(-w')(\del_{w'}g_+)(w')\right]
=(\del_{w'}h)(-w')\,,
}}
and thus $h(w') = -h(-w')+ {\rm const.}$, and similarly  for $\bar{h}(\bar{w}')$. Therefore,
\al{
h(w')+\bar{h}(-\bar{w}') =h(w') -\bar{h}(\bar{w}') + {\rm const.}
=2i {\rm Im} h(w') + {\rm const.}
}
Then, renaming $h \to h e^{i \b}$, we get 
\begin{equation}
\begin{split}
    \kappa'^2 &=  -2 \operatorname{Im} \partial_{w'} g(w') \partial_{\bar{w}'} \bar{g}(-\bar{w}') \\
    \GG' &= -2 \operatorname{Im} g(w') \bar{g}(-\bar{w}') -2 \operatorname{Im} e^{i \b}h(w') + {\rm const.} \\
     \partial_{w'} h(w') &= g(w')\partial_{w'} g(-w')- g(-w')  \partial_{w'} g(w')\,.
\end{split}
\end{equation}
%\XZ{It seems we have to make a choice for the constant to be either zero or purely imaginary. Do you agree Antoine?} \AV{I think it gets absorbed in the primitive in A.26 so it does not matter what it is. But I agree it A.20 it should be a imaginary number (inside the Im). 
%At the end I think it corresponds to the freedom of shifting the electric potential by a constant. You might worry that $\eqref{eq:sugra_solution}$ has terms that depends explicitly on $V$ and not only its derivatives. But this term is the gauge potential $C_2$ and any constant can be gauged away by using a gauge transformation.
%}
We can furthermore redefine $g \to e^{-i\b/2}g$ and the $\b$ dependence above disappears.

% \XZ{previous version: The simplest possibility is to have  $e^{-i\b(-w')}g_-(-w') = g_+(w') \equiv g(w')$ ($\b$ is \emph{not} real a priori), such that
% \al{
% (g_+(w')\bar{g}_+(-\bar{w}'))^* =  
% \bar{g}_+(\bar{w}')  g_+(-w')
% \overset{!}{=} g_-(w')\bar{g}_-(-\bar{w}')\,,
% }
% with the constraint $\bar{\b}(-\bar{w}')-\b(w') \in 2\pi \mathbb{Z}$.\XZ{explain here why $h(w') + \bar{h}(-\bar{w}')$ is imaginary.} By taking derivatives with $\partial_{w'}$ and $\partial_{\bar{w}'}$ we see that $\b$ needs to be a real constant. Therefore $g_-(-w') = e^{i \b}g_+(w') \equiv e^{i \b}g(w')$. We get that automatically $h(w') + \bar{h}(-\bar{w}')$ is imaginary, since with this identification we have $\partial_{w'}h(w')= \partial_{w'}h(-w')$ and therefore $h(w') = -h(-w')+ {\rm const.}$, and similarly  for $\bar{h}(\bar{w}')$. Then, renaming $h \to h e^{i \b}$, we get 
% \begin{equation}
% \begin{split}
%     \kappa'^2 &=  -2 \operatorname{Im} \partial_{w'} g(w') \partial_{\bar{w}'} \bar{g}(-\bar{w}') \\
%     \GG' &= -2 \operatorname{Im} g(w') \bar{g}(-\bar{w}') -2 \operatorname{Im} e^{i \b}h(w') + {\rm const.} \\
%      \partial_{w'} h(w') &\equiv g(w')\partial_{w'} g(-w')- g(-w')  \partial_{w'} g(w'),
% \end{split}
% \end{equation} 
% where we used $h(w') = - h(-w') + {\rm const.}$.}

To make the expressions dependent only on $\{w',\bar w'\}$ but not $\{-w',-\bar{w}'\}$, we write $g(w') = g_S(w') + g_A(w')$ where $g_S$ and $g_A$ are respectively symmetric and antisymmetric\footnote{This is always possible since $g(w') = \frac{1}{2} \left( g(w')+g(-w')\right)+ \frac{1}{2} \left(g(w')-g(-w') \right)$.}.
Then we get
\begin{equation}
\begin{split}
    \kappa'^2 &=  4 \operatorname{Im} \partial_{w'} g_S \,
    \partial_{\bar{w}'}\bar{g}_A \\
    \mathcal{G}' &= 4 \operatorname{Im} \left[ g_S \bar{g}_A- \frac{1}{2} h \right] + {\rm const.}, \qquad 
    \partial_{w'} h  = 2 g_A \partial_{w'} g_S-2 g_S \partial_{w'} g_A.
\end{split}
\end{equation} 
Since nothing depends explicitly on the coordinates $w'$, we can make a holomorphic transformation and treat $g_A$ as the new coordinate. We note that all dependence on $\partial_{w'} g_A$ disappears in the chain rule. Let us see this in more detail. First we have
\begin{equation}
    \begin{split}
        \kappa'^2 &
        = 4\operatorname{Im} \partial_{w'} g_S \,
    \partial_{\bar{w}'}\bar{g}_A \frac{\del_{w'}g_A}{\del_{w'}g_A}
        = 4|\partial_{w'} g_A|^2 \operatorname{Im} \frac{\partial_{w'} g_S}{\partial_{w'} g_A}, \\
        \mathcal{G}' &= 4\operatorname{Im}(g_S \bar{g}_A - \frac{h}{2}) + {\rm const.}, 
        \qquad \frac{\partial_{w'} h }{\partial_{w'} g_A} = 2 g_A \frac{\partial_{w'} g_S}{\partial_{w'} g_A}-2 g_S.
    \end{split}
\end{equation}
Next using the chain rule we have
\begin{equation}
    \frac{\del_{w'}g_S}{\del_{w'}g_A}
    % \equiv\frac{\partial g_S}{\partial g_A} 
    = \frac{\del g_S}{\del g_A}\equiv\partial_A g_S, 
    \qquad \frac{\partial_{w'} h}{\partial_{w'} g_A} = \frac{\del h}{\del g_A}  \equiv \partial_A h.
\end{equation}
There is still the dependence on $|\partial_{w'} g_A|$ in $\kappa'^2$. However, the only appearances of $\kappa'^2$ are in $f_R'$ as $\partial \mathcal{G}' \bar{\partial}\mathcal{G}'/\kappa'^2$ and the factors cancel using the chain rule, and in the metric as 
\begin{equation}
   ds^2_\Sigma \sim  \kappa'^2 dw' d\bar{w}' \sim \partial_{w'} g_A dw' \partial_{\bar{w}'} \bar{g}_A d \bar{w}' = dg_A d \bar{g}_A
\end{equation}
and they also cancel. Therefore, the solution does not depend on $\partial_{w'} g_A$. We will now simplify the notation as
\begin{equation}
    g_A \equiv z + i \rho, \qquad g_S(g_A) \to f(z+i\r), \qquad  \partial_A \to \partial, \qquad |\partial g_A|^2 \kappa'^2 \to \kappa'^2
\end{equation}
and similarly for the barred quantities. Now we can write $h$ more explicitly in terms of the primitive of $f$. We have 
\begin{equation}
    h = \int d g_A \ \partial_A h = 2\int dg_A\ \left( g_A \partial_A f - f\right) = 2 g_A f - 4 \int f.
\end{equation}
Plugging this result in $\mathcal{G}'$ we get 
%\AV{I don't think we have the 4's in $\kappa^2$ here}
\begin{equation}
    \begin{split}
        \kappa'^2 =  4\operatorname{Im} \partial f =  \partial \bar{\partial} \mathcal{G}',
        %\qquad
        %\del h = 2(z+i\r)\del f - 2 f\,,
         \qquad
        \mathcal{G}' =  -8\left(\rho \operatorname{Re}f -\operatorname{Im}\int f \right)\,,
    \end{split}
    \label{eq:kappa' and G'}
\end{equation}
where the constant in $\GG'$ is absorbed in the integral.

Let us pause and summarize our chain of redefinitions and record the relation between the original functions $\mathcal{A}_\pm$ and the new ones. We have\footnote{Here we set $\beta=0$ since we saw that it can be eliminated by further redefinitions.}
\begin{equation}
    \begin{split}
        \mathcal{A}_+(w) &= g(w') = g_S(w')+ g_A(w')= f(z,\rho) + z + i \rho, \\
        \bar{\mathcal{A}}_+(\bar{w}) &= \bar{g}(-\bar{w}') = \bar{g}_S(\bar{w}')- \bar{g}_A(\bar{w}') = \bar{f}(z,\rho) - z + i \rho, \\
        \mathcal{A}_-(w) &= g(-w') = g_S(w')- g_A(w')= f(z,\rho) - z - i \rho, \\
        \bar{\mathcal{A}}_-(\bar{w}) &= \bar{g}(\bar{w}') = \bar{g}_S(\bar{w}')+ \bar{g}_A(\bar{w}') = \bar{f}(z,\rho) + z - i \rho. \\
    \end{split}
\end{equation}
Combining this with \eqref{eq:definition of G, kappa, fR} and \eqref{eq:definiiton of del B} quickly leads to \eqref{eq:kappa' and G'}. Now the metric reads
\begin{equation}
    ds^2 =  \sqrt{6 \mathcal{G}' }  \left( f_R'^{1/2} d \Omega_6^2 + \frac{1}{9 f_R'^{3/2}}d \Omega_2^2 \right) + 4\kappa'^2 \sqrt{\frac{f_R'}{6 \mathcal{G}'}} (dz^2+d\rho^2).
\end{equation}

Now the geometry can be compactly described in the two-dimensional $(\r,z)$ plane. We want to identify the boundaries in this plane where the spheres $S^2$ and $S^6$ shrink to zero.\footnote{The boundaries in $(\r,z)$ plane do not correspond to actual boundaries of physical spacetime. For example, consider a solid cylinder described by coordinates $(r \cos\theta,r\sin\q,z)$. The axis $r=0$ is the boundary in $(r,z)$ plane where the $S^1$ shrinks but it is not the boundary of the cylinder.} For this notice that 
\begin{equation}
    R_6^3 R_2 \propto \mathcal{G}'
\end{equation}
and therefore we identify the boundary as the region where $\mathcal{G}' =0$. Writing the holomorphic function $\int f = U + i W$, we have that $W$ is a harmonic function and
\begin{equation}
    \frac{1}{8}\mathcal{G}' =W-\rho \partial_\rho W.
\end{equation}
We can now introduce $V$ through $W = -\rho V$. The harmonic condition becomes
\begin{equation}
    \nabla^2 W = 0 \implies \partial_z^2 V + \frac{2}{\rho} \partial_\rho V + \partial_\rho^2 V =0,
\end{equation}
and therefore $V$ is a harmonic function of a four-dimensional axially symmetric system, where $\rho$ is the radial variable and $z$ the height of the cylinder. The boundary condition becomes 
\begin{equation}
    \frac18\mathcal{G}'= \rho^2 \partial_\rho V = 0,
\end{equation}
and has a form suitable for the electrostatic analogy, as we derive carefully in the main text. Expressing all the supergravity fields in terms of $V$,  changing coordinates and redefining $V$ according to 
\begin{equation}
    (\rho,z) \to \frac{2}{\mu}(\rho,z), \qquad V \to \frac{2}{\mu^4}V\,,
\end{equation}
we get a solution similar to \eqref{eq:sugra_solution}, 
%\AV{added :} 
although with the wrong reality conditions. Namely the RR flux is real and the NSNS flux is imaginary. Also, following the logic of appendix \ref{app_regularity}, we find that the dilaton needs to be everywhere negative. Thus the final step leading to \eqref{eq:sugra_solution} is to use the $SL(2,\mathbb{C})$ transformation on the axi-dilaton $\tau$ in \eqref{eq:B tau F3} and the 3-forms
\al{
\tau \to \frac{a \t + b}{c \t + d}\,,\qquad  \begin{pmatrix}
        H_3 \\ dC_2
    \end{pmatrix}
    \to \begin{pmatrix}
        d & -c \\ -b & a
    \end{pmatrix} \begin{pmatrix}
        H_3 \\ dC_2
    \end{pmatrix},
}
with $a=-d=i$ and $b=c=0$ to get the correct reality conditions and change the sign of the dilaton.

\section{Regularity of the geometries}
\label{app_regularity}
In this appendix we study in detail %\JP{I believe the correct english expression is "in detail" instead of "in details"} 
the regularity conditions for the solutions \eqref{eq:sugra_solution}. We want to show that the metric and the exponential of the dilaton are positive and smooth everywhere, provided that we add a suitable background potential.

The metric is positive definite if 
\begin{equation}
     \dot{V} >0, \qquad  V''<0, \qquad  \Delta >0,
\end{equation}
and $e^\phi$ is positive if in addition 
\begin{equation}
    3 \dot{V} + \rho V'' <0\,,
\end{equation}
is  satisfied everywhere. 

To show these conditions we use properties of harmonic functions. We have that $V$ is harmonic in 4D 
\begin{equation}
    V'' + \Ddot{V} + \frac{2}{\rho} \Dot{V} = 0,
\end{equation}
and asymptotes to
\begin{equation}
    V(\rho,z) \underset{r \to \infty}{\simeq} -\eta z +\frac{\mu^5}{2^7}( z \rho^2 - z^3).
\end{equation}
We start with $-V''$, which is also harmonic in 4D. We recall the expansion near the disks and near the $z$ axis
\begin{equation}
\begin{split}
\label{eq:expansion_potential}
    &\rho\to 0 :  \qquad V(\rho,z) = f(z) - \frac{f''(z)}{6} \rho^2+ \frac{1}{120} f^{(4)}(z)\rho^4+...,\\
    &  z\to z_s : \qquad V(\rho,z) = V_s + g(\rho)(z-z_s)-\frac{2\dot{g}(\rho)+\rho \ddot{g}(\rho)}{6 \rho}(z-z_s)^3+...
\end{split}
\end{equation}
We find 
\begin{equation}
    -V'' \underset{r \to \infty}{\simeq} \frac{3 z}{64}\mu^5, \qquad -V'' \underset{\rho \to 0}{\simeq} -f''(z), \qquad -V'' \underset{z \to z_s}{\simeq} 0
\end{equation}
It is zero on the disks, regular on the $z$ axis, and is positive at infinity. Since no local minima can exist between the disks and infinity, it needs to be positive everywhere. 
%\AV{Maybe this is a bit fast, do we need to also show that $f''<0 ?$ In Maldacena's paper they only check disks and infinity and say it's enough...}

We can study $\dot{V}$ similarly. Since $\rho>0$ it is equivalent to study the positivity of $\dot{V}/\rho$, which is harmonic in 5D. We get 
\begin{equation}
    \frac{\dot{V}}{\rho} \underset{r \to \infty}{\simeq} \frac{ z}{64}\mu^5, \qquad \frac{\dot{V}}{\rho} \underset{\rho \to 0}{\simeq} - \frac{7}{6}f''(z), \qquad \frac{\dot{V}}{\rho} \underset{z \to z_s}{\simeq} 0
\end{equation}
Again we find that the function is positive at infinity, vanishes on the disks and is regular on the $z$-axis. In fact on the $z$ axis we have $\dot{V}/\rho \sim -V''>0$. Therefore we find that the function is positive everywhere.

To conclude with the positivity of the metric we want to check that $\Delta >0$. Using Laplace's equation we can write 
\begin{equation}
    \Delta =(\rho \ddot{V}-\dot{V}) (-V'')+ \rho \dot{V}'^2.
\end{equation}
Therefore it is sufficient to show $\rho \ddot{V}-\dot{V} = \rho^2 \partial_\rho( \dot{V}/\rho) >0$. Since the function $\dot{V}/\rho$ is harmonic and growing at infinity, its derivative is positive. Therefore $\Delta>0$. We conclude that the metric is positive everywhere in the region $z \geq 0$ and $\rho \geq 0$.

For the dilaton, we can use the Laplace's equation to write 
\begin{equation}
    -\rho V'' -3 \dot{V} = \rho \ddot{V}- \dot{V},
\end{equation}
which is the same function that we studied in the previous paragraph. Therefore we conclude that $e^\phi>0$.

Having determined that the metric and dilaton are positive everywhere, we still have to study in detail the smoothness of the solution. The dangerous regions are the $z-$axis and the disks, where again we can expand with \eqref{eq:expansion_potential} to look at the expressions of all supergravity fields.
Near the $z$-axis we find, up to terms of order $\rho$ or higher,
\begin{equation}
    \begin{split}
        ds^2 &\underset{\rho \to 0}{\simeq} \left(\frac{2^{12} }{45 \mu^{10} (5f'''^2-3f'' f^{(4)})^3} \right)^{1/4}\left[(5f'''^3-3f''f^{(4)}) (dz^2 + d\rho^2)+5 f''^2 \rho d\Omega_2^2 \right], \\
        e^\phi &\underset{\rho \to 0}{\simeq}  \frac{\mu^3}{\sqrt{5}}\frac{f^{(4)}}{f'' \left(5f'''^2-3f'' f^{(4)} \right)^{1/2}}, \qquad 
        \chi \underset{\rho \to 0}{\simeq} \frac{i}{\mu^3}\frac{f'' \left(5f'''^2-3f'' f^{(4)} \right)^{1/2}}{f^{(4)}} \\
          B_2 &\underset{\rho \to 0}{\simeq}  -\frac{8}{3 \mu}\left(z+ \frac{5 f'' f'''}{3f'' f^{(4)}-5f'''^3}\right) d\Omega_2, \qquad C_2 \underset{\rho \to 0}{\simeq} \frac{8 i}{3 \mu^4}\left(f+ \frac{5 (f' f'' f'''-3f''^3)}{3f'' f^{(4)}-5f'''^3}\right) d\Omega_2,
    \end{split}
\end{equation}
and we see that nothing diverges. Similarly near the disks we find at the leading order in $(z-z_s)$
\begin{equation}
    \begin{split}
        ds^2 &\underset{z \to z_s}{\simeq} \left(\frac{2^{12} \rho^2 \dot{g}^{3}}{3^3 \mu^{10} (2\dot{g}+\rho \ddot{g})} \right)^{1/4}\left[\frac{2\dot{g}+\rho \ddot{g}}{\rho f'} (dz^2 + d\rho^2)+3 \rho d\Omega_6^2 \right], \\
        e^\phi &\underset{z \to z_s}{\simeq} - \mu^3\frac{\dot{g}-\rho \ddot{g}}{\sqrt{3}\rho \left( 2\dot{g}^4+\rho \dot{g}^3 \ddot{g} \right)^{1/2}}, \qquad 
        \chi \underset{z \to z_s}{\simeq} -\frac{i}{\mu^3} \left(g+\frac{3 \rho \dot{g}^2}{\dot{g}-\rho \ddot{g}} \right) \\
          B_2 &\underset{z \to z_s}{\simeq}  -\frac{8}{3 \mu}z_s d\Omega_2, \qquad C_2 \underset{z \to z_s}{\simeq} \frac{8 i V_s}{3 \mu^4}d\Omega_2,
    \end{split}
\end{equation}
and again everything stays finite when $z \to z_s$. Therefore we determined that the solution is regular everywhere, even near $\rho=0$ and the disks.

\section{Electrostatics with a high disk}
\label{app:electrostatics}
In this appendix we solve the electrostatic problem in the limit where we have only one conducting disk at $z_s=d \gg 1$\footnote{See \cite{Copson_1947} for a solution in usual $D=3$ electrostatics.}. Our goal is to compute the constant potential on that disk and the octopole of the asymptotic electrostatic potential.

The potential of a ball of radius $a$ and charge density $\sigma(\rho)$ is
\begin{equation}
    V(\rho,z) = \frac{1}{4 \pi^2} \int d \Omega_2 \int_0^a u^2 du \frac{\sigma(u)}{(z-d)^2+\rho^2+u^2-2u \rho \operatorname{cos}\theta}.
\end{equation}
Integrating over the angles we get
\begin{equation}
    V(\rho,z) = \frac{ 1}{4\pi \rho} \int_0^a du\ u \sigma(u) \operatorname{log}\left(1+ \frac{4 \rho u}{(\rho-u)^2+(z-d)^2} \right).
    \label{potential V}
\end{equation}
In particular 
\begin{equation}
    V(\rho,d) = \frac{1}{2\pi \rho} \int_0^a du\ u \sigma(u) \operatorname{log}\left| \frac{\rho+u}{\rho-u} \right|.
\end{equation}
So we want to solve for $V_0$ in 
%\XZ{here $V_{\text{other disks}}$ is just 0, right?}
\begin{equation}
\begin{split}
    &\frac{1}{2 \pi \rho } \int_0^a du\ u \sigma(u) \operatorname{log}\left|\frac{\rho+u}{\rho-u} \right| = V_0-V_{bg}-V_{\text{other disks}} \equiv \frac{1}{2\pi \rho} f(\rho), \qquad r<a,
\end{split}
\end{equation}
where $V_{bg}(\rho)= \alpha (\rho^2 d-d^3)$ is the background potential at $z=d$.
The solution to that integral equation is \cite{integrals}
\begin{equation}
    \rho \sigma(\rho) = \frac{1}{\pi^2} \text{p.v.}\int_{-a}^a du\ \sqrt{\frac{a^2-u^2}{a^2-\rho^2}} \frac{f'(u)}{\rho-u}.
\end{equation}
In the limit $d \to \infty$ we can neglect the contributions from the image disk since it decays as $d^{-2}$. Requiring that $\sigma(a)=0$ we get the radius
\begin{equation}
    a = \sqrt{\frac{2(\alpha d^3+V_0)}{3 \alpha d}}.
\end{equation}
We still have to determine $V_0$ which can be done by computing $Q = 4 \pi \int_0^a \rho^2 \sigma(\rho)$. Using the result for $a$ we get 
\begin{equation}
    Q =  2\pi^2 \frac{(\alpha d^3+V_0)^2}{3 d \alpha},
\end{equation}
which allows to solve for $V_0$ simply by inverting 
\begin{equation}
    V_0 = -\alpha d^3 + \frac{1}{\pi}\sqrt{\frac{3}{2} Q d \alpha }.
\end{equation}
We also compute the first terms in the multipole expansion, taking into account the image disk. The distance between the two disks is $2d$. With the asymptotic expansion \eqref{eq:asymptotic_expansion} we get the first terms 
\begin{equation}
\begin{split}
    V(\rho,z) &= \frac{P}{2\pi^2} \frac{ z}{(z^2+\rho^2)^2}+ \frac{q_3}{\pi^2} \frac{z(z^2-\rho^2)}{(z^2+\rho^2)^4}+... \\
    P&= 2 Q d, \qquad q_3 = 2 Q d^3 - 8 \pi d \int_0^a du\ u^4 \sigma(u).
\end{split}
 \end{equation}
Using our result for $\sigma(\rho)$ in the limit $d \gg 1$, we can then compute 
\begin{equation}
    q_3 = 2 Qd^3 +  \sqrt{\frac{2d Q^3}{3 \alpha}} +...
\end{equation}

\bibliographystyle{utphys} % We choose the "plain" reference style
\bibliography{refs}

\end{document}